\RequirePackage{ifpdf}
\ifpdf
  \documentclass[10pt,a4paper,pdftex]{article}
\else
  \documentclass[10pt,a4paper,dvips]{article}
\fi

\ifpdf
  \usepackage{color}
\fi

\usepackage[latin1]{inputenc}
\usepackage{amsmath}
\usepackage{amsfonts}
\usepackage{amssymb}
\usepackage{float}
\usepackage{epsfig}
\usepackage{subfigure}
\usepackage[section]{placeins}
\usepackage{pstricks}
\usepackage{mathrsfs}

\ifpdf
  \usepackage{graphicx}
  \usepackage[unicode]{hyperref}
\else
  \usepackage{graphics}
\fi

\newcommand{\linkFixer}[2]{\ifpdf
			     \texorpdfstring{#1}{#2}
			   \else
			     #1
			   \fi
			  }

\newcommand{\td}{\mathrm{d}}

\newcommand{\linSpan}{span}
\newcommand{\const}{\textrm{const.}}

\newcommand{\Lie}[1]{\mathcal{#1}}
\newcommand{\superN}{\mathcal{N}}

\allowdisplaybreaks[1]

\author{Jérôme Gaillard\footnote{pyjg@swansea.ac.uk} \; and \:
  Johannes Schmude\footnote{pyjs@swansea.ac.uk}\\
  \mbox{}\\
  Department of Physics\\
  Swansea University, Swansea, SA2 8PP, United Kingdom}

\title{On the geometry of string duals with backreacting flavors}

\date{}

\begin{document}
\numberwithin{equation}{section}

\maketitle

\begin{abstract}
Making use of generalized calibrated geometry and $G$-struc\-tures we
put the problem of finding string-duals with smeared backreacting
flavor branes in a more mathematical setting.
This more formal treatment of the problem allows us to easily
smear branes without good coordinate representations, establish
constraints on the smearing form and identify a topological central
charge in the SUSY algebra. After exhibiting our methods for a series
of well known examples, we apply them to the problem of flavoring a
supergravity-dual to a $d=2+1$ dimensional $\superN=2$ super
Yang-Mills-like theory. We find new solutions to both the flavored and
unflavored systems. Interpretating these turns out to be difficult.
\end{abstract} 

\newpage

\tableofcontents

\section{Introduction}
\label{sec:introduction}

The AdS/CFT correspondence \cite{Maldacena:1997re,Witten:1998qj} gives
a by now well understood duality between type IIB string theory on
$AdS_5 \times S^5$ and maximally supersymmetric $\mathcal{N}=4$ super
Yang-Mills. This high level of supersymmetry stands in stark contrast
to the gauge theories of the standard model and so the correspondence
was quickly extended to duals of gauge theories with reduced
supersymmetry by either placing branes at singularities
\cite{Klebanov:1998hh}\cite{Klebanov:2000hb}\cite{Klebanov:2000nc} or wrapping
them on collapsing cycles
\cite{Witten:1998zw}\cite{Maldacena:2000yy}\cite{Gomis:2001aa}\cite{Gauntlett:2001ps}\cite{Bigazzi:2001aj}.

In the search for something akin to a string dual of QCD, a further
undesirable feature of the original AdS/CFT duality is the lack of
matter fields charged under the fundamental representation of the
gauge group. The issue, addressed in \cite{Karch:2002sh}, was resolved
by the inclusion of $N_f$ probe flavor branes into the background. As
these are taken to wrap non-compact cycles, their world-volume gauge
theories become infinitely weakly coupled and non dynamical from a
lower dimensional point of view. Therefore those branes provide the
global flavor symmetry group for the gauge theory. One has to assume
$N_f \ll N_c$ for the probe approximation to be valid.

However the study of some aspects of both QCD and supersymmetric
gauge theories demands for a discussion of $N_f \sim N_c$, so it is
imperative to go beyond the probe approximation. When
doing so one should keep in mind a fundamental difference between the
color and flavor branes. While the former undergo a geometric
transition and are replaced by their fluxes, the latter are still
present in the string dual -- whether one includes their backreaction
or treats them as probes -- in order to realize the $\Lie{SU}(N_f)$
flavor symmetry in the bulk. Therefore one needs to consider the
combined action
\begin{equation}
  \label{eq:general-action-of-backreacted-system}
  S = S_{IIB} + S_{\textrm{Branes}}
\end{equation}
where $S_{\textrm{Branes}} = S_{DBI} + S_{WZ}$ is the brane action
given by the DBI and Wess-Zumino terms. This method of computing the
backreaction was introduced in \cite{Klebanov:2004ya}. The 
localized branes become $\delta$-function sources in the equations of
motion, making the search for solutions of the above system highly
non-trivial. See however \cite{Burrington:2004id,Kirsch:2005uy}.

The most successful method of dealing with this issue was first
developed in \cite{Bigazzi:2005md}. See also
\cite{Casero:2006pt,Casero:2007jj,HoyosBadajoz:2008fw}. By considering
a continuous
distribution of flavor branes over their transverse directions one
avoids the problem of the inclusion of localized sources, making the
search for solutions much more feasible. An advantage of this
so-called smearing method lies in the fact that the inclusion of
localized sources breaks local isometries on the string theory side,
leading to a violation of global symmetries in the full dual gauge
theory -- including the Kaluza-Klein modes. Such
symmetries are restored after smearing. By now, many examples of
string duals with both massless and massive flavors in various
dimensions have been constructed
\cite{Benini:2006hh}\cite{Caceres:2007mu}\cite{Benini:2007gx}\cite{Casero:2007pz}\cite{Benini:2007kg}\cite{Bigazzi:2008zt}\cite{Canoura:2008at}\cite{Arean:2008az}\cite{Paredes:2006wb}\cite{Zeng:2007ta}\cite{Bigazzi:2008gd}\cite{Bigazzi:2008ie}.

Smearing of D$p$-branes is usually determined by use of a
$(10-p-1)$-form $\Omega$, the smearing form, which is in general
interpreted as a
distribution density of the branes. One issue resides in finding a way
to construct $\Omega$, such that it will be possible to find a
solution to the equations of motion. This problem is usually dealt
case by case and can be quite difficult to address when flavor
embeddings cannot be identified with globally defined coordinates. If
there is no obvious choice of $\Omega$, one uses the physical
properties of the anticipated dual gauge theory such as the mass of
the fundamental fields or the unbroken symmetries, in order to impose
constraints on its form.

In order to address this problem, we will make use of some of the
methods of modern string phenomenology. If one allows for the misnomer
of thinking of string duals with flavors as string compactifications with
non-compact internal manifolds, the two fields become virtually the
same. So it is quite striking that, while the set-up of
gauge/string duality with flavors is quite similar to that of string
phenomenology, the methods used are quite different. Especially the
advanced mathematical methods of modern phenomenology such as the uses
of generalized calibrated geometry, $G$-structures or generalized
geometry, have up to now been absent from any discussion of flavors
and duality.

Geometric arguments have been used to tackle the question of
supersymmetry conservation, but not to answer the issue of
smeared flavors. However geometry is one of the main techniques used
in the study of string compactifications, which is in many points
similar to the search for backgrounds with gauge duals.

So this paper is a first step towards bridging the gap between
string phenomenology and gauge/string dualities with flavors, by
presenting a systematic method of
finding backgrounds with smeared flavors, using tools from modern
geometry. The main goal is to find the smearing form $\Omega$. The
strategy is to use generalized calibrated geometry
\cite{Gutowski:1999tu}. The central concept of this field is that of
the calibration form $\hat{\phi}$, a $(p+1)$-form which can usually be
constructed as a bilinear of the supersymmetry spinors of the
background. It has the property that a brane is supersymmetric if and
only if the pull-back $X^* \hat{\phi}$ of the form onto the
world-volume is equal to the induced volume form. It follows
immediately that one can write the DBI action of any supersymmetric
brane in terms of the pull-back of the calibration form. As all the
backgrounds considered in this paper are type IIB backgrounds with
only the dilaton $\Phi$ and one or two Ramond-Ramond fields excited,
it will not be necessary to make use of the full machinery of
generalized calibrations. The reader interested should refer to
\cite{Koerber:2005qi,Martucci:2005ht,Koerber:2006hh} and references therein.

Let us turn to the central argument of this paper. In the case of
type IIA/B backgrounds with Ramond-Ramond flux $F_{(p+2)}$, we can write
the action of the smeared flavor branes in Einstein frame as (see
\cite{Koerber:2006hh})
\begin{equation} \label{eq:introDBI}
  S_{\textrm{Branes}} = -T_p \int_{\mathcal{M}_{10}}
  (e^{\frac{p-3}{4}\Phi} \hat{\phi} - C_{(p+1)} ) \wedge \Omega 
\end{equation}
As we will see, it is always possible to
relate the smearing form to the calibration form using supersymmetry and
the equations of motion as
\begin{equation}
  \label{eq:intro-d-star-d}
  \td \lbrack * e^{\frac{10-2p-4}{4}\Phi} \td (e^{\frac{p-3}{4}\Phi}
  \hat{\phi} ) \rbrack = \pm 2 \kappa_{10}^2 T_p \Omega
\end{equation}
giving us a geometric constraint on the smearing form. In the
following we shall study how equations (\ref{eq:introDBI}) and
(\ref{eq:intro-d-star-d}) can be applied to address the problem of smeared
flavors.

Proceeding rather pedagogically, section \ref{sec:three-examples}
introduces the methods outlined above studying three different,
well-known examples. We will see that our methods are not only capable
of reproducing the known results, yet also provide some new,
interesting ones. The examples studied are the $\mathcal{N}=1$
sQCD-like
dual of \cite{Casero:2006pt,Casero:2007jj,HoyosBadajoz:2008fw}, the
$d=2+1$ dimensional $\mathcal{N}=1$ theory of \cite{Canoura:2008at}
and the Klebanov-Witten theory \cite{Klebanov:1998hh} with massless
\cite{Benini:2006hh} and massive \cite{Bigazzi:2008zt} flavors.
Following this we shall turn to the generic case (section
\ref{sec:the-generic-case}), showing how the action
\eqref{eq:introDBI} can be constructed from purely geometric
considerations and proving its equivalence with other actions used in
the field of smeared flavors.

In section \ref{sec:gauge-string-duality-in-d=2+1} we shall finally
apply our methods to the problem of flavoring a
background dual to an $\superN = 2$ super Yang-Mills-like theory first
studied in \cite{Gomis:2001aa,Gauntlett:2001ur}. We will see that we
are able to do so without an explicit knowledge of the brane
embeddings used. We find new analytic and asymptotic solutions to the
flavored and unflavored equations of motion and discuss various
properties of these backgrounds.

Following \cite{Gauntlett:2003cy} we will show for the examples
considered, how all constraints imposed by supersymmetry upon
space-time can be understood and recovered from geometric grounds
using methods such as $G$-structures.

In the appendix \ref{sec:short-review-of-calibrations} we give a short
review of the required background in generalized calibrations. This is 
followed by a detailed example of how to calculate a calibration form
in appendix
\ref{sec:explicit-example-of-calculating-a-calibration-form}.

\section{The geometry of smeared branes}
\label{sec:geometry-of-smeared-branes}
In the following we shall now investigate what generalized calibrated
geometry can
teach us about string theory duals with backreacting, smeared flavor
branes. First we will take a detailed look at three
examples \cite{Casero:2006pt,Canoura:2008at,Benini:2006hh}.
For each of these we will briefly summarize the conventional approach
to flavoring and will then show explicitly that it can be nicely
understood in terms of a suitable calibration form. In section
\ref{sec:the-generic-case} we will turn to the case of a generic
supergravity dual.

\subsection{Three examples}
\label{sec:three-examples}

\subsubsection{\linkFixer{The string dual to an $\superN=1$ sQCD-like
    theory}{The string dual to an N=1 sQCD-like theory}}
\label{sec:examples-sQCDDual}

\paragraph{\linkFixer{Review of the $\superN=1$ sQCD-like string
    dual}{Review of the N=1 sQCD-like string dual}}

As a first example we shall turn to the string dual to an $\superN=1$
sQCD-like theory
\cite{Casero:2006pt,Casero:2007jj,HoyosBadajoz:2008fw}. It is based on
the background of \cite{Maldacena:2000yy} which
is given by the following solution of the type IIB equations of
motion:\footnote{
Except where explicitly noted, we shall always use Einstein frame in
this paper.
}
\begin{equation}
  \label{eq:maldacena-nunez-background}
  \begin{aligned}
    \td s^2 &= \alpha^\prime g_s N_c e^{\frac{\Phi}{2}} \left\lbrack
      \frac{1}{\alpha^\prime g_s N_c} \td x_{1,3}^2 + \td r^2
      + e^{2h} ( \td\theta^2 + \sin^2 \theta \td\phi^2) +
      \frac{1}{4}(\tilde{\omega}_i - A^i)^2 \right\rbrack \\
    F_{(3)} &= -\frac{1}{4} \bigwedge_a (\tilde{\omega}_a - A^a) +
    \frac{1}{4} \sum_a F^a \wedge (\tilde{\omega}_a - A^a)
  \end{aligned}
\end{equation}
with
\begin{equation}
  \label{eq:maldacena-nunez-background-elaborations}
  \begin{aligned}
    A^1 &= -a(r) \td\theta &
    \tilde{\omega}_1 &= \cos\psi \td\tilde{\theta} + \sin\psi
    \sin\tilde{\theta} \td\tilde{\phi} \\
    A^2 &= a(r) \sin\theta \td\phi &
    \tilde{\omega}_2 &= -\sin\psi \td\tilde{\theta} + \cos\psi
    \sin\tilde{\theta} \td\tilde{\phi} \\
    A^3 &= -\cos\theta \td\phi &
    \tilde{\omega}_3 &= \td\psi + \cos\tilde{\theta} \td\tilde{\phi}
  \end{aligned}
\end{equation}
The metric describes a space with topology
$\mathbb{R}^{1,3} \times \mathbb{R} \times S^2 \times S^3$, where the
three-sphere is parametrized by the
Maurer-Cartan forms $\tilde{\omega}_i$ and the one-forms $A^i$
describe the fibration between the two spheres. It is interpreted as
the near-horizon geometry of a stack of $N_c$ D5-branes wrapping an
$S^2$, thus describing the dynamics of $d=3+1$ dimensional
$\superN=1, \Lie{SU}(N_c)$ super Yang-Mills theory coupled to some
extra matter. To keep the discussion as simple as
possible, we shall focus on the so-called singular solution
which is obtained from the assumption $a(r)=0$.

The possibility of adding probe flavor branes to the above background
(\ref{eq:maldacena-nunez-background}) was studied in
\cite{Nunez:2003cf}. Using $\kappa$-symmetry the authors found several
classes of flavor D5-branes; the simplest of these is given by branes
extending along $(x^\mu, r)$ and wrapping $\psi$. They are pointlike
on the four-dimensional submanifold given by
$(\theta,\phi,\tilde{\theta},\tilde{\phi})$ and extend to $r=0$, thus
describing massless flavors. In what follows, the most important
feature of this embedding is that we are able to identify world-volume
coordinates $\xi^\alpha$ with space-time ones, $(x^\mu, r, \psi)$. So
even at the level of the space-time coordinates $X^M$ there is a very
well defined notion of coordinates tangential and transverse to the
brane.

From the perspective of type IIB string theory, it is clear that the addition of a
large number of such branes to the system
(\ref{eq:maldacena-nunez-background}) will deform the geometry of the
background. Given the form of the brane embeddings it follows that a
suitable ansatz for the deformed background should be of the
form
\begin{equation}
  \label{eq:deformed-maldacena-nunez-background}
  \begin{aligned}
    \td s^2 &= e^{2 f(r)} \lbrack \td x_{1,3}^2 + \td r^2
    + e^{2 h(r)} ( \td\theta^2 + \sin^2 \theta \td\phi^2) \\
    &+ \frac{e^{2 g(r)}}{4} (\tilde{\omega}_1^2 + \tilde{\omega}_2^2) +
    \frac{e^{2k(r)}}{4} (\tilde{\omega}_3 + \cos\theta \td\phi)^2 \rbrack \\
    F_{(3)} &= -2 N_c e^{-3f-2g-k} e^{123} +
    \frac{N_c}{2} e^{-3f-2h-k} e^{\theta\phi 3}
  \end{aligned}
\end{equation}
as the flavor branes are points on the four-dimensional transverse
manifold while singling out the $\Lie{U}(1) \subset S^3$ parametrized
by $\psi$. When writing (\ref{eq:deformed-maldacena-nunez-background})
we introduced a vielbein
\begin{equation}
  \label{eq:deformed-maldacena-nunez-vielbein}
  \begin{aligned}
    e^{x^i} &= e^f \td x^i &
    e^1 &= \frac{e^{f+g}}{2} \tilde{\omega}_1 &
    e^2 &= \frac{e^{f+g}}{2} \tilde{\omega}_2 &
    e^3 &= \frac{e^{f+k}}{2} (\tilde{\omega}_3+\cos\theta\td\phi) \\
    e^r &= e^f \td r &
    e^\theta &= e^{f+h} \td\theta &
    e^\phi &= e^{f+h} \sin\theta \td\phi
  \end{aligned}
\end{equation}
and made use of the convention $e^{A_1 \dots A_p} = e^{A_1} \wedge
\dots \wedge e^{A_p}$.

One can also interpret the ansatz
(\ref{eq:deformed-maldacena-nunez-background}) from the gauge theory
point of view. The $\Lie{U}(1)$ describes the R-symmetry of the
flavored theory, which one demands not to be broken classically by the addition of massless
flavors.

Studying the dilatino and gravitino variations of the deformed
background one obtains the projections satisfied by the SUSY spinor
$\boldsymbol{\epsilon}$,
\begin{equation}\label{eq:MNprojections}
  \begin{aligned}
    \Gamma_{r123} \boldsymbol{\epsilon} &= \boldsymbol{\epsilon} &
    \Gamma_{r\theta\phi 3} \boldsymbol{\epsilon} &= \boldsymbol{\epsilon} &
    \boldsymbol{\epsilon} &= \sigma_3 \boldsymbol{\epsilon}
  \end{aligned}
\end{equation}
as well as the BPS equations
\begin{equation}
  \label{eq:deformed-MN-BPS-equations}
  \begin{aligned}
    4f &= \Phi \\
    h^\prime &= \frac{1}{4} N_c e^{-2h-k} + \frac{1}{4} e^{-2h+k} =
    \frac{1}{2} e^{3f} F_{\theta\phi 3} + \frac{1}{4} e^{-2h+k} \\
    g^\prime &= -N_c e^{-2g-k} + e^{-2g+k} = \frac{1}{2} e^{3f}
    F_{123} + e^{-2g+k} \\
    k^\prime &= \frac{1}{4} N_c e^{-2h-k} - N_c e^{-2g-k} -
    \frac{1}{4} e^{-2h+k} - e^{-2g+k} + 2 e^{-k} \\
    &= \frac{1}{2} e^{3f} ( F_{\theta\phi 3} + F_{123} ) - \frac{1}{4}
    e^{-2h+k} - e^{-2g+k} + 2 e^{-k} \\
    \Phi^\prime &= -\frac{1}{4} N_c e^{-2h-k} + N_c e^{-2g-k} =
    -\frac{1}{2} e^{3f} ( F_{\theta\phi 3} + F_{123} )
  \end{aligned}
\end{equation}

It is a priori not obvious that the flavor branes
mentioned earlier are still supersymmetric brane embeddings for the
deformed background for arbitrary functions $g, h, k$. One therefore
has to check again that probes with world-volume directions as before, 
$\xi^\alpha = (x^\mu, r, \psi)$, still preserve all of the backgrounds
supersymmetries.

Having deformed the original background one turns to the system given
by the combined action
(\ref{eq:general-action-of-backreacted-system}). One can anticipate 
that the brane action will contribute to the energy-momentum tensor in
the Einstein equation, add a source term for the $3$-form field
strength and modify the dilaton equation by a contribution related to
the DBI action.

For the case of $N_f$ flavor branes localized at $(\theta_0,
\phi_0, \tilde{\theta}_0, \tilde{\phi}_0)$, the brane action is ($X^*$
denoting the pull-back onto the world-volume)
\begin{equation}
  \label{eq:unsmeared-dbi+wz-action-for-sqcd-dual}
  S_{\textrm{Branes}} = T_5 \sum_{N_f} \left. \left( -\int_{\mathcal{M}_6}
    \td^6 \xi e^{\frac{\Phi}{2}} \sqrt{-\hat{g}_{(6)}} +
    \int_{\mathcal{M}_6} X^* C_{(6)} \right) \right\vert_{(\theta_0,
\phi_0, \tilde{\theta}_0, \tilde{\phi}_0)}
\end{equation}
As these branes are localized in the four transverse directions, the
equations of motion will contain $\delta$-function sources, making the
search for solutions a difficult endeavour.
The idea is therefore to smoothly distribute the branes over the
transverse directions. If one assumes a transverse brane distribution
with density
\begin{equation}
  \label{eq:n=q-sQCD-smearing-form}
  \Omega = \frac{N_f}{(4\pi)^2} \sin\theta \sin\tilde{\theta}
  \td\theta \wedge \td\phi \wedge \td\tilde{\theta} \wedge
  \td\tilde{\phi}
\end{equation}
the action (\ref{eq:unsmeared-dbi+wz-action-for-sqcd-dual}) may be
generalized to
\begin{equation}
  \label{eq:smeared-dbi+wz-action-for-sqcd-dual}
  \begin{aligned}
    S_{\textrm{Branes}} &= T_5 \left( -\frac{N_f}{(4\pi)^2}
      \int_{\mathcal{M}_{10}} \td^{10} x e^{\frac{\Phi}{2}} \sin\theta
      \sin\tilde{\theta} \sqrt{-\hat{g}_{(6)}} +
      \int_{\mathcal{M}_{10}} C_{(6)} \wedge \Omega \right) \\
    &= T_5 \left( -\int_{\mathcal{M}_{10}}
      \td^{10} x e^{\frac{\Phi}{2}} \sqrt{-g_{(10)}} \vert \Omega \vert
      + \int_{\mathcal{M}_{10}} C_{(6)} \wedge \Omega \right)
  \end{aligned}
\end{equation}
where we have defined the modulus of a $p$-form $\Omega$ as
\begin{equation}
  \label{eq:definition-modulus-of-smearing-form}
  \vert \Omega \vert \equiv \sqrt{ \frac{1}{p!} \Omega_{M_1 \dots M_p}
    \Omega^{M_1 \dots M_p} }
\end{equation}
and have checked the equality of the first and second lines by explicit
calculation.

Let us take a look at how the brane action modifies the second order
equations of motion, starting with the Ramond-Ramond field
strength. Here the relevant part of the total action is
\begin{equation}
  \label{eq:Ramond-Ramond_part_of_total_action}
  \begin{aligned}
    S &= \int_{\mathcal{M}_{10}}
    - \frac{1}{2\kappa_{10}^2} \frac{e^{-\Phi}}{2} (F_{(7)} \wedge
    *F_{(7)}) + T_5 C_{(6)} \wedge \Omega
  \end{aligned}
\end{equation}
If we vary the potential $C_{(6)}$,
\begin{equation}
  \label{eq:Variation_of_Ramond-Ramond_part_of_total_action}
  \begin{aligned}
    \delta_C S &= \int_{\mathcal{M}_{10}}
    - \frac{1}{2\kappa_{10}^2} \frac{e^{-\Phi}}{2} (\td \delta C_{(6)}
    \wedge *F_{(7)} + F_{(7)} \wedge *\td\delta C_{(6)} ) + T_5 \int
    \delta C_{(6)} \wedge  \Omega \\
    &= \int_{\mathcal{M}_{10}} \delta C_{(6)} \wedge \left(
      \frac{1}{2\kappa_{10}^2} \td * e^{-\Phi} F_{(7)} + T_5 \Omega
    \right) \\
    \Rightarrow \td F_{(3)} &= 2\kappa_{10}^2 T_5 \Omega
  \end{aligned}
\end{equation}
The change in the dilaton and Einstein equations does not take such a
nice geometric form. Choosing $T_5 = \frac{1}{(2\pi)^5},
2\kappa_{10}^2 = (2\pi)^7$, the complete equations of motion are
\begin{equation}
  \label{eq:n=1-sQCD-equations-of-motion}
  \begin{aligned}
    0 &= \td F_{(3)} - (2\pi)^2 \Omega \\
    0 &= \frac{1}{\sqrt{-g_{(10)}}} \partial_\mu ( g^{\mu\nu}
    \sqrt{-g_{(10)}} \partial_\nu \Phi) - \frac{1}{12} e^\Phi
    F_{(3)}^2 - \frac{N_f}{8} e^{\frac{\Phi}{2}}
    \frac{\sqrt{-\hat{g}_{(6)}}}{\sqrt{-g_{(10)}}} \sin\theta
    \sin\tilde{\theta} \\
    0 &= R_{\mu\nu} - \frac{1}{2} g_{\mu\nu} R - \frac{1}{2}
    \left( \partial_\mu\Phi \partial_\nu\Phi - \frac{1}{2}
      g_{\mu\nu} \partial_\lambda\Phi \partial^\lambda\Phi \right) \\
    &-
    \frac{1}{12} e^\Phi \left( 3 F_{\mu\kappa\lambda}
      F_\nu^{\phantom{\nu}\kappa\lambda} - \frac{1}{2} g_{\mu\nu}
      F_{(3)}^2 \right) - T^{\textrm{flvr}}_{\mu\nu} \\
    T^{\textrm{flvr}}_{\mu\nu} &= -\frac{N_f}{4} \sin\theta
    \sin\tilde{\theta} \frac{1}{2} e^{\frac{\Phi}{2}}
    g_{\mu\alpha} g_{\nu\beta} \hat{g}_{(6)}^{\alpha\beta}
    \frac{\sqrt{-\hat{g}_{(6)}}}{\sqrt{-g_{(10)}}}
    \end{aligned}
\end{equation}

The search for solutions of (\ref{eq:n=1-sQCD-equations-of-motion}) is
simplified considerably by a powerful result due to Koerber and
Tsimpis \cite{Koerber:2007hd} who showed that any solution to the BPS
equations satisfying the modified Bianchi identity of
(\ref{eq:Variation_of_Ramond-Ramond_part_of_total_action}) solves also
the Einstein and dilaton equations and is therefore a solution of
(\ref{eq:n=1-sQCD-equations-of-motion}).

So we turn again to the issue of the BPS equations. As the brane
embeddings are supersymmetric, the projections
(\ref{eq:MNprojections}) imposed on the spinor $\boldsymbol{\epsilon}$
remain the same. However, the three-form field strength $F_{(3)}$ is
modified by the appearance of the source term in
(\ref{eq:n=1-sQCD-equations-of-motion}). To incorporate this one makes
a new ansatz for the field strength of
(\ref{eq:deformed-maldacena-nunez-background})
\begin{equation}
  \label{eq:n=1sQCD-flavored-F3}
  F_{(3)} = -\frac{N_c}{4} e^{-3f -2g -k} e^{123} - \frac{N_f -
    N_c}{4} e^{-3f-2h-k} e^{\theta\phi 3}
\end{equation}
It follows that the BPS equations (\ref{eq:deformed-MN-BPS-equations})
change to
\begin{equation}
  \label{eq:flavored-MN-BPS-equations}
  \begin{aligned}
    4f &= \Phi \\
    h^\prime &= \frac{1}{4} (N_c-N_f) e^{-2h-k} + \frac{1}{4}
    e^{-2h+k} = \frac{1}{2} e^{3f} F_{\theta\phi 3} + \frac{1}{4}
    e^{-2h+k} \\ 
    g^\prime &= -N_c e^{-2g-k} + e^{-2g+k} = \frac{1}{2} e^{3f}
    F_{123} + e^{-2g+k} \\
    k^\prime &= \frac{1}{4} (N_c-N_f) e^{-2h-k} - N_c e^{-2g-k} -
    \frac{1}{4} e^{-2h+k} - e^{-2g+k} + 2 e^{-k} \\
    &= \frac{1}{2} e^{3f} ( F_{\theta\phi 3} + F_{123} ) - \frac{1}{4}
    e^{-2h+k} - e^{-2g+k} + 2 e^{-k} \\
    \Phi^\prime &= -\frac{1}{4} (N_c-N_f) e^{-2h-k} + N_c e^{-2g-k} =
    -\frac{1}{2} e^{3f} ( F_{\theta\phi 3} + F_{123} )
  \end{aligned}
\end{equation}
It is curious to note that when written in terms of $F_{\theta\phi 3}$
and $F_{123}$ the BPS equations of the deformed and flavored systems
are the same -- see (\ref{eq:deformed-MN-BPS-equations}) and
(\ref{eq:flavored-MN-BPS-equations}). The change in the BPS equations
stems solely from the modification of the field strength. This should
not come as a surprise, as the brane embeddings are supersymmetric.
\footnote{
Whether the BPS equations are modified by the flavoring procedure is
-- to some extent -- a matter of taste. It depends on whether
one makes a sufficiently generic ansatz for the three-form field
strength to accomodate the source term. From the perspective of a
physicist who is interested in the properties of the dual gauge theory
it is more appropriate to consider the BPS equations of the flavored
and unflavored theories as different, as some phenomena such as
Seiberg duality become apparent at the level of the first-order BPS
equations \cite{Casero:2006pt}. For a mathematician on the other hand
it might be more important to think about the close link between
supersymmetry and geometry which is evident in this paper -- the fact
that the flavor branes are supersymmetric is then reflected by
the invariance of the BPS equations in terms of
$F_{ABC}$.
}

By construction $F_{(3)}$ satisfies the modified Bianchi
identity. Thus any solution of 
(\ref{eq:flavored-MN-BPS-equations}) solves the flavoring problem for
the Maldacena-N\'u\~nez background. For a discussion of these
solutions and their physical interpretation see
\cite{Casero:2006pt,Casero:2007jj,HoyosBadajoz:2008fw}.

In the above background, the generalization of the action
(\ref{eq:unsmeared-dbi+wz-action-for-sqcd-dual}) to
(\ref{eq:smeared-dbi+wz-action-for-sqcd-dual}) is fairly intuitive
and simple, because there is only one stack of flavor branes with
world-volume coordinates that can be globally identified with
space-time coordinates. However we can already anticipate the
shortcomings of 
this definition. On a technical level, the first line of \eqref{eq:smeared-dbi+wz-action-for-sqcd-dual} is inherently
dependent on the coordinate split while the second is non-linear in
the smearing form $\Omega$. From a more formal point of view it is
also unsatisfying that the formalism of
those equations treats the DBI and
Wess-Zumino contributions to the brane action on an unequal
footing. One should recall that, roughly speaking, the DBI action
defines the tree level couplings of the brane to the NS sector of the
background while the couplings to Ramond-Ramond fields are contained in
the Wess-Zumino term. A standard string theory calculation shows the
cancellation of the effects of closed strings from the two sectors on
supersymmetric branes. So it would be desirable to see an explicit
symmetry between the two terms even after smearing. Adopting once
again a more physics centered perspective we might also wonder if
there are any constraints on the choice of the smearing form. E.g.~one
should note that the smearing form does not agree with the volume form
induced on the four-cycle $(\theta, \phi, \tilde{\theta},
\tilde{\phi})$. At first
glance it might appear that there are none. After all, the
cancellations between parallel BPS branes allow us to place them at
arbitrary separations. As we will soon see, however, there are
constraints on $\Omega$ which can be traced back to the geometric
structure of the background.

\paragraph{The perspective of generalized calibrated geometry}

The properties of generalized calibrations and their relation to
supersymmetry are discussed in detail in
appendix~\ref{sec:short-review-of-calibrations}. As the backgrounds
considered in this paper are not fully generic, yet only include
dilaton and Ramond-Ramond fields in type IIB supergravity, we will not
make use of 
the most general concept of a generalized calibration. Again we refer
to \cite{Koerber:2005qi,Martucci:2005ht}. For our purposes it is
sufficient to think 
of calibrations as $(p+1)$-forms $\hat{\phi}$, such that a $p$-brane
with embedding $X^M(\xi)$ is supersymmetric if and only if it
satisfies
\begin{equation}\label{eq:calibration-form-at-BPS-bound}
  X^*\hat{\phi}_{(p+1)} = \sqrt{-\hat{g}_{(p+1)}} \td^{p+1}\xi  
\end{equation}
As discussed in the appendix, this can be understood as a simple
rephrasing of the $\kappa$-symmetry condition on the SUSY spinor
$\boldsymbol{\epsilon}$,
\begin{equation}
  \Gamma_\kappa \boldsymbol{\epsilon} = \boldsymbol{\epsilon}
\end{equation}
For us the most interesting feature of
(\ref{eq:calibration-form-at-BPS-bound}) is that when pulled back onto
the world-volume of the brane, the calibration form is equivalent to the
induced volume form, and one may write the DBI action as 
\begin{equation}
  \label{eq:DBI-in-terms-of-calibration}
  S_{DBI} = -T_{p} \int_{\mathcal{M}_{p+1}} e^{\frac{p-3}{4}\Phi}X^*\hat{\phi}
\end{equation}
Furthermore, if the $p$-brane couples electrically to the flux given
by $F_{(p+2)}$, supersymmetry in the Einstein frame requires
\cite{Gutowski:1999tu} 
\begin{equation}\label{eq:exterior-derivative-of-generic-calibration-form}
  \td (e^{\frac{p-3}{4}\Phi}\hat{\phi}) = F_{(p+2)}  
\end{equation}

In the case at hand, the calibration six-form is given by
\begin{equation}
  \label{eq:n=1-sQCD-calibration-form-definition}
  \hat{\phi} = \frac{1}{6!} (\boldsymbol{\epsilon}^\dagger \sigma_3
  \otimes \Gamma_{a_0 \dots a_5} \boldsymbol{\epsilon}) e^{a_0 \dots a_5}
\end{equation}
As explained in appendix
\ref{sec:explicit-example-of-calculating-a-calibration-form},
evaluation of the calibration form requires only the chirality of the
type IIB spinors,
$\boldsymbol{\epsilon} = \Gamma^{11} \boldsymbol{\epsilon}$
and knowledge
of the projections imposed on the SUSY spinors
(\ref{eq:MNprojections}). From the last of these it
follows that one of the Majorana-Weyl spinors of type IIB is fixed to
zero, $\boldsymbol{\epsilon} = \left( \begin{smallmatrix} \epsilon \\
    0 \end{smallmatrix}\right)$. Thus there is only one
calibration six-form and we may use $\epsilon$ instead of
$\boldsymbol{\epsilon}$. In section
\ref{sec:gauge-string-duality-in-d=2+1} we will encounter an example
with two calibration forms. Combining the SUSY projections
(\ref{eq:MNprojections}) with the definition
(\ref{eq:n=1-sQCD-calibration-form-definition}) yields
\begin{equation}
  \begin{aligned}
    \hat{\phi}_{x^0x^1x^2x^3\theta\phi} &= \epsilon^\dagger
    \Gamma_{x^0x^1x^2x^3\theta\phi} \epsilon = -\epsilon^\dagger
    \Gamma_{r 123} \epsilon = -1
  \end{aligned}
\end{equation}
The second equality makes use of chirality, the third of the SUSY
projections and the normalization $\epsilon^\dagger
\epsilon = 1$. When calculating calibration forms it is actually more
difficult to show that certain components vanish. However, the process
is rather straightforward and discussed in considerable detail in
appendix
\ref{sec:explicit-example-of-calculating-a-calibration-form}. When the
dust settles, we are left with
\begin{equation}
  \label{eq:n=1-sQCD-calibration-form}
  \begin{aligned}
    \hat{\phi} &= e^{x^0 x^1 x^2 x^3} \wedge ( e^{r3} - e^{\theta\phi} -
    e^{12} )
  \end{aligned}
\end{equation}
As $e^3$ is the only part of the vielbein containing $\td\psi$, it is
obvious that equation (\ref{eq:calibration-form-at-BPS-bound}) is
satisfied and we recover the result of \cite{Nunez:2003cf} that the
embedding in question is supersymmetric. Noting that
\begin{equation}
  \begin{aligned}
    \sqrt{-\hat{g}_{(6)}} \td^6\xi &= e^{x^0 x^1 x^2 x^3 r 3} \\
    \Omega &= 4 N_f e^{-4f -2g -2h} e^{\theta\phi 1 2}
  \end{aligned}
\end{equation}
it is easy to see that we may write the smeared brane action
(\ref{eq:smeared-dbi+wz-action-for-sqcd-dual}) as
\begin{equation}
  \label{eq:dbi+wz-action-for-sqcd-dual-using-calibration}
  \begin{aligned}
    S_{\textrm{branes}} &= T_5 \int_{\mathcal{M}_{10}}
      ( -e^{\frac{\Phi}{2}} \hat{\phi} + C_{(6)} ) \wedge \Omega
  \end{aligned}
\end{equation}
In opposite to (\ref{eq:smeared-dbi+wz-action-for-sqcd-dual}) this is
independent of coordinates, linear in the smearing form, and treats
the DBI and Wess-Zumino contributions to the brane action on an equal
footing.

Concerning the supersymmetry condition
(\ref{eq:exterior-derivative-of-generic-calibration-form}), we find
\begin{equation}
  \begin{aligned}
    \td (e^{\frac{\Phi}{2}} \hat{\phi}) = e^{-f+\frac{\Phi}{2}}
    e^{x^0 x^1 x^2 x^3} \wedge \lbrack e^{-2g} &( 2 e^k - 6 e^{2g}
    f^\prime -2e^{2g} g^\prime - e^{2g} \Phi^\prime ) e^{r12} \\
    + e^{-2h} &( \frac{1}{2}e^k - 6 e^{2h} f^\prime - 4 e^{2h}
    h^\prime - e^{2h} \Phi^\prime ) e^{r\theta\phi} \rbrack
  \end{aligned}
\end{equation}
Using the BPS equations (\ref{eq:deformed-MN-BPS-equations}) or
(\ref{eq:flavored-MN-BPS-equations}), one may verify for the
three-form field strength with (\ref{eq:n=1sQCD-flavored-F3}) and
without sources (\ref{eq:deformed-maldacena-nunez-background}) that
$\td(e^{\frac{\Phi}{2}} \hat{\phi}) = F_{(7)}$ is satisfied. We can exploit
the calibration form even further. From $e^{-\Phi} * F_{(7)} =
F_{(3)}$ and $\td F_{(3)} = (2\pi)^2 \Omega$ it follows that
\begin{equation}
  \label{eq:n=q-sQCD-relation-of-calibration-and-smearing-form}
  \begin{aligned}
    e^{-\Phi} * \td (e^{\frac{\Phi}{2}} \hat{\phi} ) &= F_{(3)} \\
    \td \lbrack e^{-\Phi} * \td (e^{\frac{\Phi}{2}} \hat{\phi} ) \rbrack &=
    (2\pi)^2 \Omega
  \end{aligned}
\end{equation}
Again note that these equations hold with or without the backreaction
of the source terms -- in the latter case with $\Omega = 0$. One
should think of them rather as a characteristic of the supersymmetries
preserved by the background than a property of the branes.

When we first introduced the smearing form in
(\ref{eq:n=q-sQCD-smearing-form}) it appeared that its choice was
rather arbitrary. After all supersymmetry allows us to place branes at
arbitrary separations. However,
(\ref{eq:n=q-sQCD-relation-of-calibration-and-smearing-form}) is not a
result of supersymmetry alone yet rather an interplay of supersymmetry
and the Einstein equations, as the following illustrates.
\begin{equation}
  \begin{aligned}
    \td (e^{\frac{\Phi}{2}}\hat{\phi}) &\overset{\textrm{SUSY}}{=} F_{(7)},
    &
    * e^{-\Phi} F_{(7)} &\equiv F_{(3)}, &
    \td F_{(3)} &\overset{\textrm{EOM}}{=} (2\pi)^2 \Omega
  \end{aligned}
\end{equation}

\paragraph{BPS equations and $G$-structures}

We showed before that the requirement of supersymmetry is related to
geometry, notably with the calibration form. As supersymmetry gives us
the BPS equations of the system, it is logical to think that one can
retrieve those equations through geometric considerations, namely
$G$-structures. When looking at the supersymmetric gravitino equation, we
can identify $F_{(3)}$ with a torsion (straightforward in string
frame), defining a new covariant derivative $\tilde{\nabla}_{\mu}$
such that
\begin{equation}
	\tilde{\nabla}_{\mu} \epsilon = \epsilon
\end{equation}
This means that we have a covariantly constant spinor satisfying
certain projections \eqref{eq:MNprojections}. $\epsilon = \sigma_3
\epsilon$ states that there is only one structure. The other two tell
us that in the six-dimensional internal manifold, there is a
covariantly constant complex chiral spinor $\eta$ verifying 
\begin{equation}
  \begin{aligned}
    \gamma_{r123} \eta &= \eta &
    \gamma_{r\theta\phi 3} \eta &= \eta
  \end{aligned}
\end{equation}
where $\gamma_i$ are the gamma matrices of the six-dimensional
internal manifold. We can choose the chirality of $\eta$ to be
\begin{equation}
  \imath \gamma_{r123 \theta \phi} \eta = - \eta
\end{equation}
Then we recognize that the six-dimensional manifold is a generalized
Calabi-Yau. It has a K\"ahler two-form $J$ and a holomorphic
three-form $\Omega$ defined as
\begin{align}
	J_{mn} &= \imath \eta^{\dagger} \gamma_{mn} \eta \\
	\Omega_{mnp} &= \eta^T \gamma_{mnp} \eta
\end{align}
Supersymmetry imposes the following conditions on the forms (see
\cite{Gauntlett:2003cy}):
\begin{align}
	\td (e^{\Phi} \ast_6 J) &= 0 \\
	\td (e^{\frac{5}{4}\Phi} \Omega) &= 0
\end{align}
From those equations, plus the generalized calibration condition
\eqref{eq:n=q-sQCD-relation-of-calibration-and-smearing-form}, we can
retrieve the BPS equations of the system, imposing $4f =
\Phi$. Indeed, this last condition, describing how the internal
manifold is embedded in space-time, cannot be captured by those
geometric properties that concern only the six-dimensional manifold. It can however easily be found using the supersymmetric variations of the dilatino and the gravitino.

\subsubsection{\linkFixer{An $\mathcal{N}=1$, $d=2+1$ example}{An N=1,
    d=2+1 example}}
\label{sec:examples-3d}

We turn now to the string dual of a $d=2+1$ dimensional
$\mathcal{N}=1$ theory that was discussed in \cite{Canoura:2008at}. We
will leave the discussion rather brief, only exhibiting the
equivalence of the actions
(\ref{eq:smeared-dbi+wz-action-for-sqcd-dual}) and
(\ref{eq:dbi+wz-action-for-sqcd-dual-using-calibration}) for this
example. In comparison to the $\mathcal{N}=1$ sQCD-like dual of
the previous section the situation is complicated by the fact that there
are three stacks of branes. While it is possible to find coordinates
such that the worldvolume of one of these stacks may be identified
with space-time coordinates, it is not possible to do so for all three
stacks simultaneously. The system has the topology
$\mathbb{R}^{1,2} \times \mathbb{R} \times S^3 \times S^3$. As in section
\ref{sec:examples-sQCDDual}, we shall work with a simplification, the
truncated system, for which the background is given by
\begin{equation}
  \label{eq:3d-unquenched-dual_background}
   \begin{aligned}
    e^{x^i} &= e^f \td x^i \quad
    e^r = e^f \td r \quad
    e^i = \frac{e^{f+h}}{2} \sigma^i \quad
    e^{\hat{i}} = \frac{e^{f+g}}{2} (\omega^i - \frac{1}{2} \sigma^i) \\
    F_{(3)} &= -2 N_c e^{-3g-3f} e^{\hat{1}\hat{2}\hat{3}}
    + \frac{1}{2} N_c e^{-g-2h-3f} (e^{13\hat{2}} - e^{12\hat{3}} -
    e^{23\hat{1}})
  \end{aligned}
\end{equation}
$\sigma^i$ and $\omega^i$ are sets of Maurer-Cartan forms
parametrizing the two three-spheres. The projections satisfied by the
SUSY spinor $\boldsymbol{\eta}$ are
\begin{equation}\label{eq:3d-SUSY-projections}
  \begin{aligned}
    \Gamma_{1\hat{1}2\hat{2}} \boldsymbol{\eta} &= -\boldsymbol{\eta} &
    \Gamma_{1\hat{1}3\hat{3}} \boldsymbol{\eta} &= -\boldsymbol{\eta} &
    \Gamma_{2\hat{2}3\hat{3}} \boldsymbol{\eta} &= -\boldsymbol{\eta} &
    \Gamma_{r\hat{1}\hat{2}\hat{3}} \boldsymbol{\eta} &= \boldsymbol{\eta} &
    \boldsymbol{\eta} &= \sigma_3 \boldsymbol{\eta}
  \end{aligned}
\end{equation}
And the BPS equations take the form
\begin{equation}
  \label{eq:3d-BPS-equations}
  \begin{aligned}
    \Phi^\prime &= N_c e^{-3g} - \frac{3}{4} N_c e^{-g-2h} \\
    h^\prime &= \frac{1}{2} e^{g-2h} + \frac{1}{2} N_c e^{-g-2h} \\
    g^\prime &= e^{-g} - \frac{1}{4} e^{g-2h} + \frac{N_c}{4}
    e^{-g-2h} - N_c e^{-3g} \\
    \Phi &= 4 f
  \end{aligned}
\end{equation}

Once more, it follows from $\boldsymbol{\eta} = \sigma_3
\boldsymbol{\eta} = \left( \begin{smallmatrix} \eta \\
    0 \end{smallmatrix} \right)$ that there is only one calibration
six-form which is given by (assuming 
$\Gamma^{11} \boldsymbol{\eta} = -\boldsymbol{\eta}$)
\begin{equation}  \label{eq:3d-unquenched-dual_calibration-form}
  \begin{aligned}
    \hat{\phi} &= e^{012} \wedge (e^{r1\hat{1}} + e^{r2\hat{2}} +
    e^{r3\hat{3}} - e^{123} + e^{3\hat{1}\hat{2}} -
    e^{2\hat{1}\hat{3}} + e^{1\hat{2}\hat{3}})
  \end{aligned}
\end{equation}
From the calibration condition for supersymmetric branes, $X^*\hat{\phi} =
\td \xi^{6} \sqrt{-\hat{g}_{(6)}}$, one can see immediately that there
are supersymmetric 5-brane embeddings 
with tangent vectors\footnote{\label{fn:integrability}
When labeling brane embeddings in terms of their tangent vectors one
should think of the brane being along the submanifold spanned by the
integral curves of the tangent vector fields. That is, if one were to
find coordinates $y^M$ such that
\begin{equation*}
  \begin{aligned}
    \partial_{x^0} &= \partial_{y^0} &
    \partial_{x^1} &= \partial_{y^1} &
    \partial_{x^2} &= \partial_{y^2} &
    E_r &= \partial_{y^3} &
    E_1 &= \partial_{y^4} &
    E_{\hat{1}} &= \partial_{y^5}
  \end{aligned}
\end{equation*}
the corresponding $1\hat{1}$ brane embedding would be given by
\begin{equation*}
  \begin{aligned}
    Y^\alpha(\xi) &= \xi^\alpha \quad &
    Y^a &= \const &
    \alpha &\in \{0,\dots, 5\} &
    a &\in \{6,\dots, 9\}
  \end{aligned}
\end{equation*}
One should note however, that it is necessary to verify, that the
distribution given by the tangent vectors is integrable, i.e.~to
verify that the coordinates $y^M$ exist. One can do so using Frobenius
theorem, which states that a distribution given by vectors $T_a$ is
integrable iff it is in involution, that is iff $[ T_a, T_b ] =
f_{abc} T_c$.
} $(\partial_{x^0}, \partial_{x^1}, \partial_{x^2},
E_r, E_i, E_{\hat{i}})$, $i \in \{ 1, 2, 3\}$. We also learn from
(\ref{eq:3d-unquenched-dual_calibration-form}) that these embeddings
are absolutely equivalent. They were
originally derived in \cite{Canoura:2008at} using
$\kappa$-symmetry. There the authors introduced a standard set of
Maurer-Cartan forms $\omega, \sigma$ to parametrize the two $S^3$s,
and then found a coordinate representation of the
$(\partial_\mu, \partial_r, E_3, E_{\hat{3}})$ branes given by
$(x^\mu,r,\psi_1,\psi_2)$. Subsequently they argued from the
symmetries of the space that there are also $1\hat{1}$ and $2\hat{2}$
embeddings, whose coordinate representation would become apparent upon
using different Maurer-Cartan forms. As we mentioned earlier, it does
not seem to be possible to find global coordinates for this system in
which all three flavor brane embeddings have good coordinate
representations -- thus this is an ideal setting for using the
calibration form (\ref{eq:3d-unquenched-dual_calibration-form}).

Our analysis here shall start with the $3\hat{3}$
embeddings. In \cite{Canoura:2008at} their smeared action was given by
\begin{equation}  \label{eq:3d-unquenched-dual_smeard-action}
  \begin{aligned}
    S_{D5} &= T_5 \left( -\int \td^{10}x e^{\frac{\Phi}{2}}
      \sqrt{-G_{10}} \vert \Omega^{(1)} \vert +
      \int_{\mathcal{M}_{10}} C_{(6)} \wedge \Omega^{(1)} \right) \\
    \Omega^{(1)} &= -\frac{N_f}{\pi^2} e^{-4f-2h-2g}
    e^{12\hat{1}\hat{2}} \\
    \vert \Omega^{(1)} \vert &= \frac{N_f}{\pi^2} e^{-4f-2h-2g} \\
    \sqrt{-G_{10}} &= \frac{1}{64} e^{10f+3g+3h} \sin\theta
    \sin\tilde{\theta} \\
    \sqrt{-\hat{G}_6} &= \frac{1}{4} e^{6f+g+h}
  \end{aligned}
\end{equation}
Now
\begin{equation}
  \begin{aligned}
    \hat{\phi} \wedge \Omega^{(1)} &= -\frac{N_f}{\pi^2} e^{-4f-2h-2g}
    \sqrt{-G_{10}} \td^{10}x = \td^{10}x \sqrt{-G_{10}} \vert
    \Omega^{(1)} \vert
  \end{aligned}
\end{equation}
Thus again, we may write the action of one stack of $(3\hat{3})$
branes as
\begin{equation}
  \begin{aligned}
    S_{D5} &= T_5 \int_{\mathcal{M}_{10}} \left( -e^{\frac{\Phi}{2}}
        \hat{\phi} + C_{(6)} \right) \wedge \Omega^{(1)}
  \end{aligned}
\end{equation}
The above may be easily generalized to the case of three stacks of
D5-branes as the expression is linear in $\Omega$.
\begin{equation}\label{eq:3d-dual-smeared-calibrated-brane-action}
  \begin{aligned}
    S_{D5} &= T_5 \int_{\mathcal{M}_{10}} \left( -e^{\frac{\Phi}{2}}
      \hat{\phi} + C_{(6)} \right) \wedge \Omega \\
    \Omega &= \Omega^{(1)} + \Omega^{(2)} + \Omega^{(3)} \\
    \Omega^{(2)} &= -\frac{N_f}{\pi^2} e^{-4f-2h-2g}
    e^{13\hat{1}\hat{3}} \\
    \Omega^{(3)} &= -\frac{N_f}{\pi^2} e^{-4f-2h-2g}
    e^{23\hat{2}\hat{3}}
  \end{aligned}
\end{equation}
Where $\Omega^{(2)}$ is the smearing form for branes extending along
$2\hat{2}$ and $\Omega^{(3)}$ smears the $1\hat{1}$ embedding. The
linearity of the above expression gives a good motivation for the use
of $\sum_i \vert \Omega^{(i)} \vert$ instead of $\vert \Omega \vert$
in the original action of \cite{Canoura:2008at}
\begin{equation}\label{eq:3d-dual-smeared-oldschool-brane-action}
  S_{D5} = T_5 \left( -\int \td^{10}x e^{\frac{\Phi}{2}}
    \sqrt{-G_{10}} \sum_{i=1}^3 \vert \Omega^{(i)} \vert +
    \int_{\mathcal{M}_{10}} C_{(6)} \wedge \Omega \right)
\end{equation}

Independently of whether one uses the action
(\ref{eq:3d-dual-smeared-calibrated-brane-action}) or
(\ref{eq:3d-dual-smeared-oldschool-brane-action}) the Bianchi identity
is modified to $\td F_{(3)} = -2\kappa_{10}^2 T_5 \Omega$ -- the minus
sign being due to the convention $e^\Phi F_{(3)} = - * F_{(7)}$ used
in \cite{Canoura:2008at}. Accordingly one changes the ansatz for the
field-strength by adding a term $f_{(3)}$ which is not closed,
\begin{equation}
  \label{eq:3d-flavored-field-strength}
  \begin{aligned}
    F_{(3)} &\mapsto F_{(3)} + f_{(3)} \\
    f_{(3)} &= 2 N_f e^{-g-2h-3f} ( e^{12\hat{3}} + e^{23\hat{1}} -
      e^{13\hat{2}} )
  \end{aligned}
\end{equation}
The BPS equations (\ref{eq:3d-BPS-equations}) change to
\begin{equation}
  \label{eq:3d-BPS-equations-flavored}
  \begin{aligned}
    \Phi^\prime &= N_c e^{-3g} - \frac{3}{4} (N_c - N_f) e^{-g-2h}
    \\
    h^\prime &= \frac{e^{g-2h}}{2} + \frac{N_c-4N_f}{2} e^{-g-2h} \\
    g^\prime &= e^{-g} - \frac{1}{4} e^{g-2h} - N_c e^{-3g} +
    \frac{N_c-4N_F}{4} e^{-g-2h}
    \\
    \Phi &= 4 f    
  \end{aligned}
\end{equation}

Let us now turn to the SUSY condition
(\ref{eq:exterior-derivative-of-generic-calibration-form}). A
straightforward yet tedious calculation yields
\begin{equation}
  \label{eq:d=3-exterior-differential-of-calibration}
  \begin{aligned}
    \td (e^{\frac{\Phi}{2}}\hat{\phi}) &= e^{\frac{\Phi}{2}-f} e^{012}
    \wedge \{
    (2 e^{-g}- 6f^\prime - 2 g^\prime - h^\prime - \Phi^\prime)
    (e^{r1\hat{2}\hat{3}} - e^{r2\hat{1}\hat{3}} +
    e^{r3\hat{1}\hat{2}}) \\
    &+ \frac{e^{-2h}}{2} (-3 e^g + 12 e^{2h} f^\prime + 6
    e^{2h} h^\prime + e^{2h} \Phi^\prime) e^{r123} \}
  \end{aligned}
\end{equation}
Using the BPS equations (\ref{eq:3d-BPS-equations}) or
(\ref{eq:3d-BPS-equations-flavored}) respectively one can verify that
$-e^{-\Phi} * \td (e^{\frac{\Phi}{2}}\hat{\phi}) = F_{(3)}$ is satisfied in
both the deformed and flavored case. Furthermore we know that $\td
F_{(3)} = (2\pi)^2 \Omega$, thus we are again able to obtain a
constraint on the smearing form as
\begin{equation}
  \label{eq:3d-d-Star-d-thing}
  \begin{aligned}
    (2\pi)^2 \Omega &= \td \lbrack -e^{-\Phi} * \td
    (e^{\frac{\Phi}{2}}\hat{\phi}) \rbrack 
  \end{aligned}
\end{equation}
We immediately see why there have to be three stacks of flavor
D5-branes in the backreacted solution -- the calibration form respects
the symmetries of the two three-spheres and from
(\ref{eq:3d-d-Star-d-thing}) it follows that the same holds true for
the smearing form. It would therefore not be possible to obtain a
smeared system with only one or two of the three stacks.

We can again use $G$-structures to derive the BPS equations for the
system. In this case the internal manifold is seven-dimensional, with
a covariantly constant spinor which satisfies
\begin{equation}
  \begin{aligned}
    \gamma_{1\hat{1}2\hat{2}} \eta &= -\eta &
    \gamma_{1\hat{1}3\hat{3}} \eta &= -\eta &
    \gamma_{r\hat{1}\hat{2}\hat{3}} \eta &= \eta
  \end{aligned}
\end{equation}
We recognize here a generalized $G_2$ holonomy manifold with the
associative three-form $\hat{\phi}$ defined as
\begin{equation}
	\hat{\phi}_{mnp} = -\imath \bar{\eta} \gamma_{mnp} \eta
\end{equation}
The condition imposed by supersymmetry is
\begin{equation}
	\td (e^{\Phi} \ast_7 \hat{\phi}) = 0
\end{equation}
Together with the generalized calibration condition, and assuming
$\Phi = 4f$, this condition provides us with a method to rederive the BPS
equations (\ref{eq:3d-BPS-equations}),
(\ref{eq:3d-BPS-equations-flavored}).

\subsubsection{The Klebanov-Witten model}
\label{sec:examples-Klebanov-Witten}

Finally we take a look at the Klebanov-Witten model for the cases of
massless \cite{Benini:2006hh} and massive flavors
\cite{Bigazzi:2008zt}. The Klebanov-Witten model
\cite{Klebanov:1998hh} is based on D3-branes at the tip of the
conifold and is dual to a certain $\mathcal{N}=1$ super Yang-Mills
theory. So apart from the dilaton and the metric there is self-dual
$F_{(5)}$ flux due to the D3s. In contrast to the previous two
examples, one uses D7s to introduce flavor degrees of freedom into the
system. These source $F_{(1)}$, so the suitable ansatz for the
relevant deformed, flavored background is
\begin{equation}
  \label{eq:deformed-kw-background}
  \begin{aligned}
    \td s^2 &= h^{-\frac{1}{2}} \td x_{1,3}^2 \\
    &+ h^{\frac{1}{2}} \left\lbrack e^{2f}
      \td\rho^2 + \frac{e^{2g}}{6} \sum_{i=1,2} (\td\theta_i^2 +
      \sin^2\theta_i \td\phi_i^2) + \frac{e^{2f}}{9} (\td\psi +
      \sum_{i=1,2} \cos\theta_i \td\phi_i)^2 \right\rbrack \\
    F_{(5)} &= 27 \pi N_c e^{-4g-f} h^{-5/4} (
    e^{x^0 x^1 x^2 x^3 \rho} - e^{\theta_1 \phi_1 \theta_2 \phi_2
      \psi}) \\
    F_{(1)} &= \frac{N_f(\rho)}{4\pi} (\td\psi + \cos\theta_1 \td\phi_1
    + \cos\theta_2 \td\phi_2 )
  \end{aligned}
\end{equation}
with $\psi \in \lbrack 0,4\pi \rbrack, \theta_i \in \lbrack
0,\pi\rbrack, \phi_i \in \lbrack 0,2\pi\rbrack, \rho \in
\mathbb{R}$. There is an obvious choice of vielbein
\begin{equation}
  \label{eq:kw-vielbein}
  \begin{aligned}
    e^{x^i} &= h^{-1/4} \td x^i &
    e^\rho &= h^{1/4} e^f \td\rho \\
    e^{\theta_i} &= \frac{1}{\sqrt{6}} h^{1/4} e^g \td\theta_i &
    e^{\phi_i} &= \frac{1}{\sqrt{6}} h^{1/4} e^g \sin\theta_i
    \td\phi_i \\
    e^\psi &= \frac{1}{3} h^{1/4} e^f (\td\psi + \cos\theta_1
    \td\phi_1 + \cos\theta_2 \td\phi_2) &
  \end{aligned}
\end{equation}

The flavor branes behave differently in the massless or massive
case. In the former, the authors of \cite{Benini:2006hh} used two
stacks of branes whose world-volume coordinates may once more be
identified with space-time ones,
\begin{equation}\label{eq:kw-massless-embedding}
  \begin{aligned}
    \xi_1^\alpha &= ( x^\mu, \rho, \theta_2, \phi_2, \psi ) & \theta_1
    &= \textrm{const.} & \phi_1 &= \textrm{const.} \\
    \xi_2^\alpha &= ( x^\mu, \rho, \theta_1, \phi_1, \psi ) & \theta_2
    &= \textrm{const.} & \phi_2 &= \textrm{const.}
  \end{aligned}
\end{equation}
So prior to smearing the system has a global $\Lie{U}(N_f) \times
\Lie{U}(N_f)$ flavor symmetry -- one for each set of D7s. This is
obviously a four-parameter family of embeddings, which can be smeared
over the transverse $(\theta_i, \phi_i)$ directions. In the
massive case the embeddings are more complicated. In the field theory,
the mass term breaks the global symmetry to the diagonal $\Lie{U}(N_f) \times
\Lie{U}(N_f) \mapsto \Lie{U}(N_f)$, which corresponds the two stacks
joining into one on the string theory side. There is again a 
four-parameter family of brane embeddings, yet as the generic
embedding is much more complicated than those
of~\eqref{eq:kw-massless-embedding}, we shall only look at one
representative, trusting that the calibration form
will ensure that we make use of the whole family of branes. Choosing
world-volume coordinates
$\xi = (x^\mu, \theta_1, \phi_1, \theta_2, \phi_2)$, this is given by
\begin{equation}
  \label{eq:kw-massive-embedding}
  \begin{aligned}
    X^M(\xi) &= \left( x^\mu, \rho_q - \frac{2}{3} \log \sin
      \frac{\theta_1}{2} - \frac{2}{3} \log \sin \frac{\theta_2}{2},
      \theta_1, \phi_1, \theta_2, \phi_2, \phi_1 + \phi_2 + 2\beta
    \right) \\
    \rho_q, \beta &= \const
  \end{aligned}
\end{equation}
The constant $\rho_q$ denotes the minimal radius reached by the brane
and may therefore be identified as the mass.

The branes have an $(7+1)$-dimensional world-volume and we therefore
need to construct the calibration $8$-form. In the case at hand this
requires the knowledge of the supersymmetric spinors on the
conifold. These were discussed in \cite{Arean:2004mm}. Our conventions
however are those of \cite{Benini:2006hh}. The SUSY spinor $\epsilon$
is related  to a constant spinor
$\eta$ as $\epsilon = h^{-1/8} e^{-\frac{\imath}{2}\psi} \eta$. Both
satisfy the projections
\begin{equation}
  \label{eq:kw-SUSY-projections}
  \begin{aligned}
    \imath \sigma_2 \otimes \Gamma_{x^0 x^1 x^2 x^3} \eta &= \eta &
    \Gamma_{r\psi} &= \imath \sigma_2 \eta \\
    \Gamma_{\theta_1\phi_1} &= -\imath \sigma_2 \eta &
    \Gamma_{\theta_2\phi_2} &= -\imath \sigma_2 \eta
  \end{aligned}
\end{equation}
From equation (\ref{eq:definition-of-calibration-form}) it follows
that the calibration form for D7-branes is given by
\begin{equation}\label{eq:kw-definition-of-calibration}
  \hat{\phi} = \frac{1}{8!} (\eta^\dagger \imath \sigma_2 \otimes
  \Gamma_{a_0 \dots a_7} \eta) e^{a_0 \dots a_7}  
\end{equation}
which we may evaluate using \eqref{eq:kw-SUSY-projections} to
be
\begin{equation}
  \label{eq:kw-calibration-form}
  \begin{aligned}
    \hat{\phi} &= e^{x^0 x^1 x^2 x^3} \wedge \left( e^{\rho \theta_1 \phi_1
        \psi} + e^{\rho \theta_2 \phi_2 \psi} - e^{\theta_1 \phi_1
        \theta_2 \phi_2} \right)
  \end{aligned}
\end{equation}
At this point we may calculate the pull-backs $X^* \hat{\phi}$ for both
embeddings \eqref{eq:kw-massless-embedding} and
\eqref{eq:kw-massive-embedding}. Finding $X^* \hat{\phi} =
\sqrt{-\hat{g}_{(8)}} \td^8 \xi$ we do thus verify that the brane
embeddings are indeed supersymmetric.

In Einstein frame, the integrand of the DBI action is
$e^{\Phi}\sqrt{-\hat{g}_{(8)}} \td^8 \xi = e^\Phi X^* \hat{\phi}$. As before,
supersymmetry requires this to satisfy $\td (e^\Phi \hat{\phi}) =
F_{(9)}$. Making use of the definition $F_{(1)} = -e^{-2\Phi}*F_{(9)}$
and the equation of motion $\td F_{(1)} = -\Omega$, we arrive at the
following
\begin{equation}
  \label{eq:kw-d-star-d-and-all-that}
  \begin{aligned}
    F_{(9)} &= \td ( e^\Phi \hat{\phi} ) = 3 h^{-\frac{1}{4}} e^{-f}
    \frac{N_f(\rho)}{4\pi} e^{x^0 x^1 x^2 x^3 \rho \theta_1 \phi_1
      \theta_2 \phi_2} \\
    F_{(1)} &= -e^{-2\Phi}*F_{(9)} = - 3 h^{-\frac{1}{4}} e^{-f}
    \frac{N_f(\rho)}{4\pi} e^\psi \\ 
    \Omega &= - \td F_{(1)} 
    = \frac{N_f(\rho)}{4\pi} (\sin\theta_1 \td\theta_1\wedge\td\phi_1 +
    \sin\theta_2 \td\theta_2\wedge\td\phi_2) \\
    &+ \frac{N_f^\prime(\rho)}{4\pi} \td\rho \wedge
    (\td\psi + \cos\theta_1 \td\phi_1 + \cos\theta_2 \td\phi_2) \\
    N_f(\rho) &= \frac{4\pi}{3} e^{-2g-\Phi} (4 e^{2g} g^\prime + e^{2g}
    \Phi^\prime - 4 e^{2f})
  \end{aligned}
\end{equation}
The name for the function $N_f(\rho)$ has been chosen in anticipation
of what is to come -- it will denote the effective number of flavors
at a given energy scale. It should not be confused with $N_f$, the
number of flavor branes.

One should notice that the only assumptions made in
deriving (\ref{eq:kw-d-star-d-and-all-that}) are the form of $F_{(5)}$
and the vielbein describing the deformed background
(\ref{eq:kw-vielbein}). That is, the above relations hold for all
types of D7-branes one might want to smear, massless or
massive. They allow us to write down the BPS equations of the system
which can be derived from the SUSY variations \cite{Bigazzi:2008zt} or
using geometric methods.
\begin{equation}
  \label{eq:kw-BPS-equations}
  \begin{aligned}
    g^\prime &= e^{2f-2g} &
    f^\prime &= 3 - 2 e^{2f-2g} - \frac{3 N_f(\rho)}{8\pi} e^\Phi \\
    \Phi^\prime &= \frac{3 N_f(\rho)}{4\pi} e^\Phi &
    h^\prime &= -27\pi N_c e^{-4g}
  \end{aligned}
\end{equation}
Note that there are four first-order equations for the five functions
$\Phi, f, g, h, N_f$. Furthermore, the smearing procedure always uses
the same action,
\begin{equation}\label{eq:kw-smeared-action}
  \begin{aligned}
    S_{\textrm{Branes}} &= T_7 \int_{\mathcal{M}_{10}} \left( -e^\Phi
      \hat{\phi} + C_{(8)} \right) \wedge \Omega
  \end{aligned}
\end{equation}
The authors of \cite{Benini:2006hh,Bigazzi:2008zt} used an action of
the type encountered in (\ref{eq:smeared-dbi+wz-action-for-sqcd-dual})
and (\ref{eq:3d-dual-smeared-oldschool-brane-action}), yet once more
the equivalence with (\ref{eq:kw-smeared-action}) may be shown
explicitly -- we will also present a general proof of the validity of
(\ref{eq:kw-smeared-action}) in section \ref{sec:the-generic-case}.

Given that the discussion up to this point is completely independent
of the type of brane one wants to smear, one might ask how to
distinguish between the different classes of potential flavor
branes. The answer to that question lies in the choice of the function
$N_f(\rho)$.

However, even before looking at specific choices of $N_f(\rho)$ the
generic form of $\Omega$ in (\ref{eq:kw-d-star-d-and-all-that}) tells
us quite a bit about possible smeared-brane configurations. For once,
it is not possible to break the $\Lie{SU}(2) \times \Lie{SU}(2) \times
\Lie{U}(1) \times \mathbb{Z}_2$ symmetry of the background, as this is
the inherent symmetry of $\Omega$ (The $\mathbb{Z}_2$ describes the
exchange of the two spheres). So for massless branes we will only be
able to smear both stacks simultaneously.

The massless branes may be identified with the coordinates
given by~\eqref{eq:kw-massless-embedding}. Thus they are smeared by
the terms proportional to $\td\theta_i\wedge\td\phi_i$. As the
smearing form is symmetric under the exchange $(\theta_1,\phi_1)
\leftrightarrow (\theta_2,\phi_2)$ it is clear that we will have to
smear both stacks of branes. I.e.~one cannot assume
$\Omega_{\theta_1\phi_1}$ to vanish without $\Omega_{\theta_2\phi_2}$
vanishing as well. The term involving $\td\rho$ on the other hand is 
not transverse to the world-volume defined
by~\eqref{eq:kw-massless-embedding}. In order to smear only massless
branes, one needs this term to vanish. I.e.~massless branes require
\begin{equation}\label{eq:kw-constraint-for-massless-branes}
  N_f^\prime(\rho) = 0
\end{equation}
Using this constraint the system (\ref{eq:kw-BPS-equations}) is fully
determined and can be solved. In that case, we can see from
\eqref{eq:kw-smeared-action} that the last term in
\eqref{eq:kw-calibration-form} -- which does not contain $e^{\rho}$ --
does not contribute. Interpreting the smearing form as a
brane-density, we may identify the overall factor with the number of
flavors,
\begin{equation}\label{eq:kw-second-constraint-for-massless-branes}
  N_f = 4\pi N_f(\rho)
\end{equation}
That is, our decision to smear $N_f$ massless branes with a constant
number of flavors imposes two constraints into the system, namely
\eqref{eq:kw-constraint-for-massless-branes} and
\eqref{eq:kw-second-constraint-for-massless-branes}.

Our choice for $N_f(\rho)$ may also be interpreted using the local
geometry of the brane embeddings instead of their global
coordinates. The vectors
\begin{equation}
  \label{eq:kw-massless-embeddings-tangent-vectors}
  (\partial_{x^\mu}, \partial_\rho, \partial_\psi)
\end{equation}
are tangent to either stack of branes. As the smearing form should --
locally -- define a volume orthogonal to these
vectors, we demand\footnote{
Interior multiplication of forms with vectors is defined as
\begin{equation*}
  (\imath_X \omega)_{N_1 \dots N_{p-1}} = X^M \omega_{M N_1 \dots N_{p-1}}  
\end{equation*}
}
\begin{equation}
  \imath_{\partial_{x^\mu}} \Omega = \imath_{\partial_\rho} \Omega =
  \imath_{\partial_\psi} \Omega = 0    
\end{equation}
It follows that $4\pi N_f(\rho) = \const = N_f$.

Turning to the massive case, the authors of \cite{Bigazzi:2008zt} used
\begin{equation}
  \label{eq:kw-massive-nf}
  N_f^\prime(\rho) = 3 N_f e^{3\rho_q - 3\rho} (3\rho-3\rho_q)
\end{equation}
In principle one would expect that one can combine the knowledge of
the embedding (\ref{eq:kw-massive-embedding}) together with the
general form for $\Omega$ in order to derive this form for
$N_f(\rho)$, as we did for massless branes, yet we were unable to do
so. The reason might be that the authors of \cite{Bigazzi:2008zt}
considered not just the single representative of the family of massive
embeddings, yet a complete distribution. Our analysis contributes to
the construction of $N_f(\rho)$ in so far, however, as the derivation
in \cite{Bigazzi:2008zt} requires the assumption that the $\Lie{SU}(2)
\times \Lie{SU}(2) \times \Lie{U}(1) \times \mathbb{Z}_2$ symmetry
cannot be broken, while we have shown that this is not an assumption,
but an innate property of the background.

Once more one invokes \cite{Koerber:2007hd} and needs only to study
the BPS equations (\ref{eq:kw-BPS-equations}) together with the
modified Bianchi identity to find solutions of the second order
equations. We refer to the original papers for a discussion of the
solutions.

Anticipating the possibility of using the formalism presented up to
this point in order to smear branes whose coordinate representation is
unknown, we shall now discuss the problem of correctly interpreting the
smearing form $\Omega$. Using the vielbein it takes the form
\begin{equation}
  \begin{aligned}
    \Omega &= \frac{6 N_f(\rho)}{\sqrt{h}} e^{-2g} (e^{\theta_1\phi_1}
    + e^{\theta_2\phi_2}) + \frac{6 N_f^\prime(\rho)}{\sqrt{h}}
    e^{-2f} e^{\rho\psi}
  \end{aligned}
\end{equation}
In the case for the massless embeddings
(\ref{eq:kw-massless-embedding}) the second term disappeared and it is
straightforward to interpret the first as a distribution on the space
transverse to the two stacks of D7s. If we did not know about the
massive embeddings (\ref{eq:kw-massive-embedding}) it would be
tempting to interpret the term including $N_f^\prime$ as the
distribution of a third stack of branes extending along $x^\mu$,
wrapping $(\theta_1, \phi_1, \theta_2, \phi_2)$ and positioned at
fixed $(\rho, \psi)$. That is we would think of this term as a
contribution of compact, smeared D7 branes. The presence of such
branes is potentially disastrous as the gauge theory in their
world-volume could remain dynamic from a four-dimensional point of
view. In the case at hand, the eight-dimensional gauge coupling
behaves as $g_{\textrm{YM}} \sim g_s \alpha^{\prime 2}$, which
vanishes for $\alpha^\prime \to 0$, the decoupling limit of the
D3s. When using D5 branes on the other hand this does not have to
happen. For the massive Klebanov-Witten model we know that our
interpretation in terms of compact D7 branes is wrong as we are
smearing a single stack of massive ones. Keeping this in mind we
conclude that it is not straightforward to know which branes have been
smeared by simply investigating $\Omega$.

Again we would like to carry on the procedure we used
previously to find the BPS equations (\ref{eq:kw-BPS-equations})
through geometric properties. However, in the previous examples, the
starting point was
to identify $F_{(3)}$ with a torsion. In the Klebanov-Witten model,
there is no $F_{(3)}$ but instead $F_{(1)}$ and $F_{(5)}$. As a
consequence, it is not straightforward to transform the supersymmetric
gravitino variation into a covariant derivative. In this case as in
the other ones, supersymmetry should nevertheless impose conditions on
the geometry of the internal manifold. Understanding how to derive
those conditions is left for future work.

\subsection{The generic case}
\label{sec:the-generic-case}

The three examples of the previous section provide us with all the
intuition needed to understand the relation between generalized
calibrated geometry and supergravity duals with backreacted, smeared
flavors. For a type IIA/B background with Ramond-Ramond flux
$F_{(p+2)}$ and arbitrary dilaton we expect that we should always be
able to write the action in terms of the calibration and smearing form
as
\begin{equation}
  \label{eq:general-smeared-brane-action}
  S_{\textrm{Branes}} = -T_p \int_{\mathcal{M}_{10}} (
  e^{\frac{p-3}{4}\Phi} \hat{\phi} - C_{(p+1)} ) \wedge \Omega
\end{equation}
Now as we discussed in appendix
\ref{sec:short-review-of-calibrations}, supersymmetry imposes
\begin{equation}
  \label{eq:supersymmetry-of-smearing-form}
  \td ( e^{\frac{p-3}{4}\Phi} \hat{\phi} ) = F_{(p+2)}
\end{equation}
Combining this with the modified $p$-form equation of motion $\td
F_{(10-p-2)} = 2\kappa_{10}^2 T_p \Omega$, as derived in
(\ref{eq:Variation_of_Ramond-Ramond_part_of_total_action}), we may link 
the calibration and the smearing form
\begin{equation}
  \label{eq:link-of-calibration-and-smearing-form}
  \td \lbrack * e^{\frac{10-2p-4}{4}\Phi} \td ( e^{\frac{p-3}{4}\Phi}
  \hat{\phi} ) \rbrack = \pm 2\kappa_{10}^2 T_p \Omega
\end{equation}
The overall sign depends on the conventions used when relating the
field strength $F_{(p+2)}$ to its dual. In what follows, we shall
give a more formal argument why the action
(\ref{eq:general-smeared-brane-action}) is appropriate to describe
smeared branes, show that it is equivalent to the actions previously used
in the literature and finally examine some of the consequences of
the above relations.

\subsubsection{The smeared brane action}
\label{sec:derivation-of-smeared-brane-action}

The problem of smearing a generic DBI+Wess-Zumino system takes a
rather simple form from a mathematical point of view. Here we are
dealing with two spaces, the world-volume $\mathcal{M}_{p+1}$ and
space-time $\mathcal{M}_{10}$, which are related by the embedding map
\begin{equation}
  \label{eq:embedding_map}
  \begin{aligned}
    X: \mathcal{M}_{p+1} &\to \mathcal{M}_{10} \\
    \xi^\alpha &\mapsto X^M(\xi)
  \end{aligned}
\end{equation}
As integrals of scalars are ill-defined on manifolds, it is mandatory
for this discussion to think of the brane action as an integral of
differential forms. For the Wess-Zumino term, the integrand
is the pull-back of the relevant electrically coupled gauge-potential
onto the world-volume, $\int_{\mathcal{M}_{p+1}} X^*
C_{(p+1)}$. Whereas we integrate over the induced volume form and the
dilaton in the case of the DBI action,\footnote{The discussion in this section
  considers branes without world-volume gauge fields or the $NS$
  potential $B$. See however \cite{Benini:2007kg,Koerber:2005qi,Martucci:2005ht}.}
$\int_{\mathcal{M}_{p+1}} \td^{p+1}\xi e^{\frac{p-3}{4}\Phi} \sqrt{-\hat{g}_{(p+1)}}$.
The crucial point is that there is no way to a priori identify the
DBI integrand with a $(p+1)$-form in space-time, as the induced volume
form is usually not thought of as the pull-back of a differential
form. Indeed, we were rather careless in section
\ref{sec:three-examples} as we did not discriminate between
the set of form-fields in the world-volume of the brane,
$\Omega (\mathcal{M}_{p+1})$, and that defined on all of
space-time, $\Omega (\mathcal{M}_{10})$.

One might argue that we should be able to somehow push the
induced volume form forward onto space-time. This is certainly the case
if we are able to identify world-volume with space-time
coordinates. In the case of the string dual of the $\mathcal{N}=1$
sQCD-like theory
this was strikingly obvious. As a matter of fact, the action written
in the first line of (\ref{eq:smeared-dbi+wz-action-for-sqcd-dual}) is
exactly of the form (\ref{eq:general-smeared-brane-action}). In a
generic situation however, we cannot expect to be able to find such a
set of global coordinates. Moreover the natural operations induced by
maps between manifolds are push-forwards of vectors and pull-backs of
forms. And as they connect spaces of different dimensions, they cannot
be assumed to be invertible.

This is where calibrated geometry comes in. As we have seen before,
supersymmetric branes satisfy $X^*\hat{\phi} = \sqrt{-\hat{g}_{(p+1)}}
\td^{p+1}\xi$. Making use of this fact allows us to treat the DBI and
Wess-Zumino terms on a democratic footing, as both integrands can now
be written as pull-backs of $(p+1)$-forms defined on space-time.

We shall now show that the action
(\ref{eq:general-smeared-brane-action}) can always be written in the
form used in \cite{Benini:2006hh,Canoura:2008at}. Essentially the
whole discussion boils down to the fact that we may locally choose
nice coordinates. Let us assume that we have a single stack of
supersymmetric $p$-branes. Locally, we may choose coordinates
$x^M = (z^\mu, y^m)$ such that the branes extend along the $z^\mu$;
that is for world-sheet coordinates $\xi^\mu$ and embeddings
$X^M(\xi)$ we have
\begin{equation}
  \partial_\nu X^M = \left\{ \begin{matrix} \delta_\nu^M & M \in
      \{0,\dots , p\} \\ 0 & M \notin \{0,\dots , p\} \end{matrix}
  \right.
\end{equation}
The vectors $\partial_\mu$ are tangent to the brane. They span a
subset of $T\mathcal{M}_{10}$ which may be thought of as the embedding
of the tangent space $T\mathcal{M}_{p+1}$ of the brane into that of
space-time. Orthonormalizing the $\partial_\mu$ we obtain a new basis
of $T\mathcal{M}_{p+1}$ given by some $E_\alpha$. I.e.~$\linSpan
(E_\alpha) = T\mathcal{M}_{p+1} \subset T\mathcal{M}_{10}$. It follows from the
construction that the $E_\alpha$ are closed under the Lie bracket,
i.e.~$\lbrack E_\alpha, E_\beta \rbrack \in \linSpan
(E_\gamma)$. Therefore $E_\alpha^m = 0$ and the matrix $E_\alpha^\mu$
is invertible. We may complete the set $E_\alpha$ to a basis of the
whole tangent space, $E_A = (E_\alpha, E_a)$. Naturally, there is a
dual basis of covectors, $e^A = (e^\alpha, e^a)$ which we may use as a
vielbein.

Having constructed a vielbein suitable for our purposes we shall now
express the DBI action in terms of that vielbein. As the two bases are
dual we have
\begin{equation}
  0 = E_\alpha e^b = E_\alpha^M e_M^b
\end{equation}
Contracting with $(E_\alpha^\mu)^{-1} = e^\alpha_\mu$, we obtain
\begin{equation}
  e_\mu^b = 0  
\end{equation}
This is quite important. It means that the components $e^a$ of the
vielbein are not pulled back onto the brane world-volume whereas all
the $e^\alpha$ are. After all,
the pull-back acts as $X^* (\omega_M \td x^M) = \omega_\mu \td
\xi^\mu$. It follows that the volume form induced onto the brane
world-volume is given by the pull-back of the forms $e^\alpha$
\begin{equation}
  \sqrt{-\hat{g}_{(p+1)}} \td^{p+1}\xi = \bigwedge_{\alpha} (X^*e^\alpha)
\end{equation}
The DBI action in this frame is therefore given by
\begin{equation}
  \label{eq:DBI-action-in-suitable-frame}
  S_{DBI} = -T_p \int_{\mathcal{M}_{p+1}} e^{\frac{p-3}{4}\Phi}
  \bigwedge_{\alpha} (X^*e^\alpha)
\end{equation}

In the final part of our discussion, we will impose some constraints
on the calibration and smearing form, and show that an action of the
form (\ref{eq:introDBI}) can always be rewritten in the form
(\ref{eq:smeared-dbi+wz-action-for-sqcd-dual}). For the calibration
form to satisfy $X^*\hat{\phi} = \sqrt{-\hat{g}_{(p+1)}} \td^{p+1}\xi$, it
has to include $\bigwedge_{\alpha} e^\alpha$. So we may assume it to
be of the form $\hat{\phi} = \bigwedge_{\alpha} e^\alpha + \tilde{\phi}$,
where $\tilde{\phi}$ is a $(p+1)$-form which does not depend on all
the indices $\alpha$ simultaneously and therefore includes some of the
$e^a$. It follows that $X^*\tilde{\phi} = 0$. The smearing form is
defined on the space transverse to the branes. This space has a
one-form basis given by $\td y^m$. As we saw above $e^a_\mu = 0$ and
it follows that we may write the smearing form in
this basis as
\begin{equation}
  \begin{aligned}
    \Omega &= \frac{1}{(10-p-1)!}
    \Omega_{m_1 \dots  m_{10-p-1}} \td y^{1} \wedge \dots \wedge \td
    y^{10-p-1} \\
    &= \frac{1}{(10-p-1)!}\Omega_{a_1 \dots a_{10-p-1}} e^{a_1
      \dots a_{10-p-1}} = \Omega_{(p+2) \dots 9} e^{(p+2) \dots 9}
  \end{aligned}
\end{equation}
That is, locally the smearing form is defined by a single sca\-lar
function $\Omega_{(p+2) \dots 9}$ and includes the wedge product over
all the transverse components of the vielbein, $\bigwedge_a e^a$. We
see immediatly that $\tilde{\phi} \wedge \Omega = 0$. Moreover
\begin{equation}
  \begin{aligned}
    \hat{\phi} \wedge \Omega &= e^{0 \dots 9} \Omega_{(p+2) \dots 9}
  \end{aligned}
\end{equation}
The trick is now to associate the indices of the function
$\Omega_{(p+2) \dots 9}$ with something other than those of the
relevant components of the vielbein, as we need those for the overall
volume form $e^{0 \dots 9} = \sqrt{-g_{(10)}} \td^{10}x$. As the form
reduces to a function and we are working in flat indices, we may
resolve this as follows:
\begin{equation}
  \begin{aligned}
    \hat{\phi} \wedge \Omega &= e^{0 \dots 9} \Omega_{(p+2) \dots 9} = e^{0
      \dots 9} \sqrt{ \Omega_{(p+2) \dots 9} \Omega^{(p+2 \dots 9)} }
    \\
    &= \sqrt{-g_{(10)}} \td^{10}x \vert \Omega \vert
  \end{aligned}
\end{equation}
with the modulus of the smearing form defined as in
(\ref{eq:definition-modulus-of-smearing-form}). As the wedge product
is linear, one may immediately generalize our argument here for
multiple stacks of branes, thus proving our initial assertion.

As an immediate application of the results of this section we shall
take a brief look at central extensions of SUSY algebras.
From the equations of motion (\ref{eq:n=1-sQCD-equations-of-motion})
it follows that the smearing form is exact, $\td F_{(10-p-2)} = 2
\kappa_{10}^2 T_p \Omega$. Supersymmetry requires that
$( e^{\frac{p-3}{4}\Phi} \hat{\phi} - C_{(p+1)} )$ is closed. It follows
that we may write the smeared brane action
(\ref{eq:general-smeared-brane-action}) as a surface integral at
infinity,
\begin{equation}
  \label{eq:general-smeared-brane-action-as-surface-integral}
  S_{\textrm{Branes}} = -\frac{1}{2\kappa_{10}^2} \int_{S^9_\infty} (
  e^{\frac{p-3}{4}\Phi} \hat{\phi} - C_{(p+1)} ) \wedge F_{(10-p-2)}
\end{equation}
This takes the form of a charge. From the original discussion of
generalized calibrated geometry in \cite{Gutowski:1999tu} we recall
the fact that probe-brane actions relate to central charges in
supersymmetry algebras -- as one would expect for BPS objects. We
conjecture that the charge defined by
(\ref{eq:general-smeared-brane-action-as-surface-integral}) has the
same interpretation.

\section{\linkFixer{$\superN=2$ gauge-string duality in $d=2+1$}{N=2
    gauge-string duality in d=2+1}}
\label{sec:gauge-string-duality-in-d=2+1}

Let us now apply the methods described in the previous section to the
flavoring of an $\superN=2$ super Yang-Mills-like dual in $d=2+1$. A
string dual can be found in the unflavored case by constructing a
domain-wall solution in $d=7$ gauged supergravity and then lift it to
ten dimensions. It then describes a stack of NS5-branes wrapping a
three-sphere. Details and physical interpretation of this solution can
be found in \cite{Gomis:2001aa} and \cite{Gauntlett:2001ur}. We are
first going to describe the unflavored solution using notations from
\cite{Gauntlett:2001ur} before studying the addition of flavors.

\subsection{The unflavored solution}

In the unflavored case, we consider only NS5-branes wrapping a three-sphere. So the non-zero fields in type IIB supergravity are the metric $g_{\mu \nu}$, the dilaton $\Phi$ and the NS-NS 3-form field strength $H$. The solution found in \cite{Gauntlett:2001ur} is, in string frame
\begin{align}
  \td s^2 &= \, \td \xi_{1,2}^2 + \frac{2z}{g^2} \td \Omega_3^2 +
  \frac{e^{2x}}{g^2} (\td z^2 + \td \psi^2) + \frac{1}{g^2 \Omega}
  \sin^2 \psi (E_1^2 + E_2^2) \\ 
  e^{2 \Phi} &= \left( \frac{2z}{g^2} \right)^{3/2}
  \frac{e^{-2A+x}}{\Omega} \\ 
  H &= \frac{g e^{-2x}}{2z \Omega^{1/2}} \left[ \cos \psi
    (e^{124}-e^{236}-e^{135}) - e^{2x} \sin \psi e^{127} \right]
  \notag \\ 
  &-\frac{g e^{-2x} \sin \psi}{\Omega^{3/2}} \left[e^{6x} \sin^2 \psi
    + e^{2x}(4 \cos^2 \psi +1) - 3e^{-2x} \cos^2 \psi - \frac{\cos^2
      \psi}{z} \right] e^{567} \notag \\ 
  &-\frac{g e^{-2x} \cos \psi}{\Omega^{3/2}} \left[e^{4x} \sin^2 \psi
    -3 + e^{-4x} \cos^2 \psi - \frac{e^{2x} \sin^2 \psi}{z} \right]
  e^{456}
\end{align}
$A$ and $x$ are functions of $z$ defined as
\begin{align}
	e^{-2x} = &\frac{I_{3/4}(z) - c K_{3/4}(z)}{I_{-1/4}(z) + c K_{1/4}(z)} \\
	e^{A+3x/2} = &z \left( I_{-1/4}(z) + c K_{1/4}(z) \right)
\end{align}
where $I_{\alpha}$ and $K_{\alpha}$ are the modified Bessel functions and $c$ is an integration constant.
In the previous equations, we used the vielbein
\begin{equation} \label{eq:Gauntlettvielbeins}
	\begin{aligned}
		e^a &= \frac{\sqrt{2z}}{g} S^a \quad a=1,2,3 & e^7 &= \frac{1}{g \Omega^{1/2}} (\cos \psi \td z - e^{2x} \sin \psi \td \psi) \\
		e^4 &= \frac{1}{g \Omega^{1/2}} (e^{2x} \sin \psi \td z + \cos \psi \td \psi) & e^8 &= \td \xi^1 \\
		e^5 &= \frac{1}{g \Omega^{1/2}} \sin \psi E_1 & e^9 &= \td \xi^2 \\
		e^6 &= \frac{1}{g \Omega^{1/2}} \sin \psi E_2 & e^0 &= \td \xi^0
	\end{aligned}
\end{equation}
with
\begin{equation}
	\begin{aligned}
		\sigma^1 &= \cos \tilde{\beta} \td \tilde{\theta} + \sin \tilde{\beta} \sin \tilde{\theta} \td \tilde{\phi} \\
   \sigma^2 &= \sin \tilde{\beta} \td \tilde{\theta} - \cos \tilde{\beta} \sin \tilde{\theta} \td \tilde{\phi} \\
   \sigma^3 &= \td \tilde{\beta} + \cos \tilde{\theta} \td \tilde{\phi} \\
		S^1 &= \cos \phi \frac{\sigma^1}{2} - \sin \phi \frac{\sigma^2}{2} \\
		S^2 &= \sin \theta \frac{\sigma^3}{2} -\cos \theta \left(\sin \phi \frac{\sigma^1}{2} + \cos \phi \frac{\sigma^2}{2} \right) \\
		S^3 &= -\cos \theta \frac{\sigma^3}{2} -\sin \theta \left(\sin \phi \frac{\sigma^1}{2} + \cos \phi \frac{\sigma^2}{2} \right) \\
		E_1 &=\td \theta + \cos \phi \frac{\sigma^1}{2} - \sin \phi \frac{\sigma^2}{2} \\
		E_2 &= \sin \theta \left(\td \phi + \frac{\sigma^3}{2} \right) - \cos \theta \left(\sin \phi \frac{\sigma^1}{2} + \cos \phi \frac{\sigma^2}{2} \right) \\
		\Omega &= e^{2x} \sin^2 \psi + e^{-2x} \cos^2 \psi \\
		\theta, &\tilde{\theta}, \psi \in [0,\pi] \qquad \phi, \tilde{\phi} \in [0,2\pi[ \qquad \tilde{\beta} \in ]0,4\pi]
	\end{aligned}
\end{equation}
and $\td \Omega_3^2= \sigma^i \sigma^i$. We know that type IIB supergravity contains thirty-two supercharges that can be described by an $\Lie{SO}(2)$ doublet of chiral spinors $\boldsymbol{\epsilon}=(\epsilon^- , \epsilon^+)$. Their chirality is expressed as
\begin{equation}
	\Gamma_{11} \boldsymbol{\epsilon} = \Gamma_{1234567890} \boldsymbol{\epsilon} = -\boldsymbol{\epsilon}
\end{equation}
This background preserves four supercharges, corresponding to $\superN=2$ in $d=2+1$ dimensions. This means that $\boldsymbol{\epsilon}$ has to verify the projections
\begin{equation} \label{eq:projections}
	\begin{aligned}
		\Gamma^{1256} \boldsymbol{\epsilon} &= \boldsymbol{\epsilon} \\
		\Gamma^{1346} \boldsymbol{\epsilon} &= \boldsymbol{\epsilon} \\
		\Gamma^{4567} \boldsymbol{\epsilon} &= \sigma_3 \boldsymbol{\epsilon}
	\end{aligned}
\end{equation}
where $\sigma_3$ is the third Pauli matrix.

\subsection{Deformation of the solution}

We are now working again in Einstein frame. We first notice that, in
the solution of the previous section, $e^4$ and $e^7$ are mixing the
$z$ and $\psi$ coordinates. In order to simplify this, we make a
common change of coordinates, first proposed in
\cite{DiVecchia:2002ks}:
\begin{equation}
	\begin{aligned}
		\rho &= \sin \psi \frac{e^{A-x/2}}{(2z g^2)^{1/4}} \\
		\sigma &= \sqrt{g} \frac{\cos \psi}{(2z)^{3/4}} e^{A+3x/2}
	\end{aligned}
\end{equation}
We then get that $e^4 = h_1(\rho,\sigma) \td \rho$ and $e^7 = h_2(\rho,\sigma) \td \sigma$. Let us now deform the metric by modifying the vielbein in \eqref{eq:Gauntlettvielbeins}
\begin{equation}
	\begin{aligned}
		e^a &= e^{-f/2} \sqrt{j(\rho,\sigma)} S^a \quad a=1,2,3 & e^7 &= e^{-f/2} \sqrt{h_2(\rho,\sigma)} \td \sigma \\
		e^4 &= e^{-f/2} \sqrt{h_1(\rho,\sigma)} \td \rho & e^8 &= e^{-f/2} \td \xi^1 \\
		e^5 &= e^{-f/2} \sqrt{h_1(\rho,\sigma) k(\rho,\sigma)} E_1 & e^9 &= e^{-f/2} \td \xi^2 \\
		e^6 &= e^{-f/2} \sqrt{h_1(\rho,\sigma) k(\rho,\sigma)} E_2 & e^0 &= e^{-f/2} \td \xi^0
	\end{aligned}
\end{equation}
It gives us the following ansatz for the metric:
\begin{equation}
	\begin{aligned}
		\td s^2 &= e^{-f(\rho,\sigma)} \Big( \td \xi_{1,2}^2 + j(\rho,\sigma) \td \Omega_3^2 + h_1(\rho,\sigma) [\td \rho^2 + k(\rho,\sigma) (E_1^2 + E_2^2)] \\
		&\qquad \qquad + h_2(\rho,\sigma) \td \sigma^2 \Big) \\
	\end{aligned}
\end{equation}
It is straightforward to see that this ansatz leaves the topology of the previous solution invariant.

\subsection{Calibration, smearing and G-structures}

We are now interested in adding flavor D5-branes to the
background. Following the usual method, we first deform the unflavored
solution for D5-branes. Then we find calibrated cycles where we can
put supersymmetric D5-branes. We finally smear them and find a
solution that includes their backreaction.

The solution in the previous section describes NS5-branes. As we are interested in the IR behaviour of the gauge dual, we want to consider D5-branes. So we first perform an S-duality on the solution. It gives a new solution of type IIB supergravity describing D5-branes, for which non-zero fields are the metric, the dilaton and the Ramond-Ramond 3-form such that
\begin{align}
	g_{\mu \nu}^{NS5} &\rightarrow g_{\mu \nu}^{D5} \\
	\Phi^{NS5} &\rightarrow -\Phi^{D5} \\
	H_{(3)}^{NS5} &\rightarrow F_{(3)}^{D5} \\
	\sigma_3 &\rightarrow \sigma_1
\end{align}

As we want to keep the same number of supercharges, and just deform the previous solution, we are imposing the same projections on the SUSY spinors as \eqref{eq:projections}. We then define a new $\Lie{SO}(2)$ doublet
\begin{equation}
	\boldsymbol{\eta} =
	\begin{pmatrix}
		\eta^- \\ \eta^+
	\end{pmatrix}
	=
	\begin{pmatrix}
		\epsilon^- + \epsilon^+ \\
		\epsilon^- - \epsilon^+
	\end{pmatrix} 
\end{equation}
such that \eqref{eq:projections} becomes
\begin{equation} \label{eq:Gprojections}
	\begin{aligned}
		\Gamma^{1256} \boldsymbol{\eta} &= \boldsymbol{\eta} \\
		\Gamma^{1346} \boldsymbol{\eta} &= \boldsymbol{\eta} \\
		\Gamma^{4567} \boldsymbol{\eta} &= \sigma_3 \boldsymbol{\eta}
	\end{aligned}
\end{equation}
Notice that $\boldsymbol{\eta}$ is still a doublet of chiral spinors that satisfies
\begin{equation}
	\Gamma_{11} \boldsymbol{\eta} = -\boldsymbol{\eta}
\end{equation}
From the third projection, we see that $\eta^-$ and $\eta^+$ are both non-zero, but behave differently under the action of gamma matrices. So for each spinor we can construct a six-dimensional generalized calibration form
\begin{equation}
	\begin{aligned}
		\hat{\Phi}^- &= \eta^{-T} \Gamma_{089abc} \eta^- e^{089abc} \\
		\hat{\Phi}^+ &= \eta^{+T} \Gamma_{089abc} \eta^+ e^{089abc} \\
	\end{aligned}
\end{equation}
Those forms can be written as
\begin{equation}
	\begin{aligned}
		\hat{\Phi}^- &= e^{089} \wedge \hat{\phi}^- \\
		\hat{\Phi}^+ &= e^{089} \wedge \hat{\phi}^+  \\
	\end{aligned}
\end{equation}
where $\hat{\phi}^+$ and $\hat{\phi}^-$ are three-forms. Using supersymmetric variations of the gravitino and the dilatino and identifying $F_{(3)}$ with a torsion term, it is possible to define two covariant derivatives $\tilde{\nabla}^+$ and $\tilde{\nabla}^-$ such that
\begin{equation}
	\begin{aligned}
		\tilde{\nabla}^+ \eta^+ &= 0 \\
		\tilde{\nabla}^- \eta^- &= 0
	\end{aligned}
\end{equation}
So the existence of $\eta^{\pm}$ imposes that the internal manifold has special
holonomy, and thus admits a corresponding $G$-structure. With each
spinor satisfying the projections \eqref{eq:Gprojections}, it is
possible to define two different $G_2$ structures in the
seven-dimensional space with tangent directions \{1,2,3,4,5,6,7\}. The
corresponding associative three-forms are $\hat{\phi}^+$ and $\hat{\phi}^-$. We
want the flavor branes we add to preserve the same supercharges as in
the unflavored solution. From \cite{Gauntlett:2003cy}, we know that
there is in fact an $\Lie{SU}(3)$ structure in that space, for which
the three-dimensional calibration form is
\begin{equation}
	\hat{\phi} = \frac{1}{2} (\hat{\phi}^- - \hat{\phi}^+) 
\end{equation}
So the calibration form for D5-branes in this geometry is
\begin{equation}
	\hat{\Phi} = e^{089} \wedge \hat{\phi}
\end{equation}
We have (details of the calculation can be found in Appendix \ref{sec:explicit-example-of-calculating-a-calibration-form})
\begin{equation}
	\begin{aligned}
		\hat{\phi}^- &= e^{123} + e^{145} - e^{167} + e^{246} + e^{257} + e^{347} - e^{356} \\
		\hat{\phi}^+ &= - e^{123} - e^{145} - e^{167} - e^{246} + e^{257} + e^{347} + e^{356}
	\end{aligned}
\end{equation}
So,
\begin{equation} \label{eq:calibration}
	\hat{\Phi} = e^{089} \wedge (e^{123} + e^{145} + e^{246} - e^{356})
\end{equation}

In order to find solutions for the deformed background, we first need to provide an ansatz for the Ramond-Ramond form $F_{(3)}$:
\begin{equation}
	\begin{aligned}
		F &= e^{-3\Phi/4} (F_{124}(\rho,\sigma) e^{124} + F_{135}(\rho,\sigma) e^{135} + F_{236}(\rho,\sigma) e^{236} + F_{127}(\rho,\sigma) e^{127} \\ 
		&\qquad + F_{456}(\rho,\sigma) e^{456} + F_{567}(\rho,\sigma) e^{567}) \\ 
	\end{aligned}
\end{equation}
and we assume the dilaton depends only on $\rho$ and $\sigma$. As mentioned previously, we know from \cite{Koerber:2007hd} that conservation of supersymmetry gives us first order differential equations that, in addition to imposing the Bianchi identity for $F_{(3)}$, will solve the equations of motion. One way to find those equations is to study the type IIB supersymmetry transformations of the dilatino and the gravitino
\begin{align}
	\delta \lambda &= \frac{1}{2} \Gamma^{\mu} \partial_{\mu} \Phi \boldsymbol{\eta} + \frac{1}{24} e^{\Phi/2} F_{\mu \nu \rho} \Gamma^{\mu \nu \rho} \sigma_3 \boldsymbol{\eta} = 0 \\
	\delta \psi_{\mu} &= \nabla_{\mu} \boldsymbol{\eta} + \frac{1}{96} e^{\Phi/2} F_{\nu \rho \sigma} (\Gamma_{\mu}^{\phantom{\mu} \nu \rho \sigma} - 9 \delta_{\mu}^{\nu} \Gamma^{\rho \sigma}) \sigma_3 \boldsymbol{\eta} = 0
\end{align}
Another way is to use the geometric properties of the space, using $G$-structures and generalized calibration conditions. As stated previously, we need to assume that $\Phi = 2f$. Otherwise, we can look at the dilatino variation to get an additional condition. From it we get
\begin{align}
	\partial_{\rho} \Phi &= \frac{e^{(2f-\Phi)/4} \sqrt{h_1}}{2} (F_{127} - F_{567}) \\
	\partial_{\sigma} \Phi &= \frac{e^{(2f-\Phi)/4} \sqrt{h_2}}{2} (F_{135} + F_{236} + F_{456} - F_{124})
\end{align}
Then we remember that $\hat{\Phi}^-$ is a generalized calibration and $\hat{\phi}^-$ defines a $G_2$ structure. So we get two conditions on those forms
\begin{align}
	\td (e^{\Phi/2} \hat{\Phi}^-) &= - e^{\Phi} \ast_{10} F \\
	\td (e^{\Phi} \ast_7 \hat{\phi}^-) &= \td (e^{\Phi} \ast_{10}
        \hat{\Phi}^-) = 0
\end{align}
Using the conditions on the dilaton, those two equations give us
\begin{align}
	f &= \frac{\Phi}{2} \label{eq:firstBPS}\\
	\partial_{\rho} \Phi &= -\frac{j \sqrt{h_1} F_{567} + h_1 \sqrt{k}}{2j} \label{eq:eqrPhi} \\
	\partial_{\sigma} \Phi &= \frac{\sqrt{h_2} (F_{456}-3 F_{124})}{2} \\
	\partial_{\rho} j &= 2 h_1 \sqrt{k} \\
	\partial_{\sigma} j &= 2j \sqrt{h_2} F_{124} \\
	\partial_{\rho} k &= 2 \sqrt{k} - \frac{h_1 k^{3/2}}{j} + k \frac{h_1^{3/2} F_{567} - \partial_{\rho} h_1}{h_1}\\
	\partial_{\sigma} k &= 0 \\
	\partial_{\rho} h_2 &= h_2 \frac{j \sqrt{h_1} F_{567} + h_1 \sqrt{k}}{j} \\
	\partial_{\sigma} h_1 &= h_1 \sqrt{h_2} (F_{124}-F_{456}) \label{eq:eqsh1} \\
	F_{127} &= - \frac{\sqrt{h_1 k}}{j} \\
	F_{135} &= - F_{124} \\
	F_{236} &= - F_{124}
\end{align}
Moreover, we must have
\begin{equation}
	\begin{aligned}
		\partial_{\rho} \partial_{\sigma} \Phi &= \partial_{\sigma} \partial_{\rho} \Phi \\
		\partial_{\rho} \partial_{\sigma} j &= \partial_{\sigma} \partial_{\rho} j
	\end{aligned}
\end{equation}
So we get
\begin{align}
	\partial_{\rho} F_{124} &= - \frac{j \sqrt{h_1} F_{124} F_{567} + h_1 \sqrt{k} (3 F_{124} + 2 F_{456})}{2j} \\
	\frac{\partial_{\rho} F_{456}}{\sqrt{h_1}} &= - \frac{\partial_{\sigma} F_{567}}{\sqrt{h_2}} - \frac{ \sqrt{h_1 k} (4 F_{124} + 5 F_{456}) + j F_{124} F_{567}}{2j} \label{eq:lastBPS}
\end{align}
Let us now eliminate components of $F$ in \eqref{eq:eqrPhi} to \eqref{eq:eqsh1} and try to solve those equations. We get
\begin{align}
	h_1 &= \frac{e^{-2\Phi}}{j} e^{a(\rho)}  \label{eq:newBPS1}\\
	h_2 &= e^{-2\Phi} e^{b(\sigma)} \\
	e^{2\Phi} &= \frac{2 \sqrt{k}}{j \partial_{\rho} j} e^a \\
	F_{124} &= \frac{e^{(a - b)/2} k^{1/4} \partial_{\sigma} j} {\sqrt{2 \partial_{\rho} j} j^{3/2}} \\
	F_{456} &= \frac{e^{(a - b)/2} k^{1/4} ( \partial_{\sigma}j \partial_{\rho}j - 2 j \partial_{\sigma} \partial_{\rho} j) }{\sqrt{2} (j \partial_{\rho}j)^{3/2}} \\
	F_{567} &= \frac{\sqrt{k} (\partial_{\rho}j)^2 - j ((2+\sqrt{k} a') \partial_{\rho}j - 2 \sqrt{k} \partial_{\rho}^2 j)}{\sqrt{2} k^{1/4} j (\partial_{\rho}j)^{3/2}} \\
	\partial_{\rho} k &= 2 \sqrt{k} - k a' \label{eq:newBPS7}
\end{align}
We notice that $b(\sigma)$ is arbitrary, which corresponds to the fact that it is always possible to redefine the $\sigma$ coordinate. To simplify the problem, we are taking $b = 0$ in the following sections.

\subsection{Addition and smearing of flavor branes}

In order to add and smear flavor branes, one needs to find the
smearing form $\Omega$. Following the prescription presented in the
first part of this article, we know that this form is related to the
calibration form of our background $\hat{\Phi}$ (see
\eqref{eq:calibration}) through
\begin{equation}
	\Omega = \td F = - \td (e^{-\Phi} \ast \td (e^{\Phi/2} \hat{\Phi})) 
\end{equation}
Using this, the ansatz for the metric and for $F$ and the equations found in the previous section (\eqref{eq:firstBPS} to \eqref{eq:lastBPS}), we can deduce that the most general form of $\Omega$ is
\begin{equation}
	\Omega = e^{\Phi} \Big( N_{f1}(\rho,\sigma) [e^{2367} + e^{1357} - e^{1247}] + N_{f2}(\rho,\sigma) e^{4567} \Big)
\end{equation}
with
\begin{align}
	\partial_{\sigma} F_{124} &= \sqrt{h_2} \frac{j (F_{124} F_{456} - 5 F_{124}^2 + 2 N_{f1} e^{2\Phi}) - 2 - 2 \sqrt{h_1 k} F_{567}}{2j} \\
	\frac{\partial_{\sigma} F_{456}}{\sqrt{h_2}} &= \frac{\partial_{\rho} F_{567}}{\sqrt{h_1}} + \frac{3 F_{567}^2}{2} + \frac{F_{567} (4j - h_1 k)}{2j \sqrt{h_1 k}} + \frac{3 F_{456} (F_{456} - F_{124})}{2} - e^{2\Phi} N_{f2}	
\end{align}
Consistency between those equations and \eqref{eq:newBPS1} to \eqref{eq:newBPS7} imposes that
\begin{align}
	N_{f2} &= N_{f1} + \frac{j}{h_1 \sqrt{k}} \partial_{\rho} N_{f1} \label{eq:eqNf} \\
	0 &= 2 j^2 \partial^2_{\rho} j + 2 e^a j \partial^2_{\sigma} j + j (\partial_{\rho} j)^2 - e^a (\partial_{\sigma} j)^2 - j^2 (a' \partial_{\rho} j + 4 e^a N_{f1}) \label{eq:j-equation}
\end{align}
We now see that the only unknown  we have is $N_{f1}$. Any function of
$\rho$ and $\sigma$ is possible and will give first order differential
equations that will solve the modified equations of motion for type
IIB supergravity plus flavor embeddings. Finding a solution then
consists only on solving the second-order differential equation
\eqref{eq:j-equation}. However, while the choice of the function $N_{f1}$
determines which branes are smeared, we are unable to derive the
embedding of the supersymmetric branes that have been smeared. One
might want to recall the discussion at the end of section
\ref{sec:examples-Klebanov-Witten}.

\subsubsection{Different possibilities for the smearing form}

As it was stated before, the starting point of adding smeared flavors is to choose a smearing form, which, in the case we are currently studying, corresponds to choosing a function $N_{f1}(\rho,\sigma)$.

A first possibility would be to take $N_{f1}$ independent of $\rho$. It follows from \eqref{eq:eqNf} that
\begin{equation}
	N_{f1} = N_{f2} = N_f(\sigma)
\end{equation}
Then we can try to solve \eqref{eq:j-equation} by making the following ansatz for $j$:
\begin{equation}
	j(\rho,\sigma) = G(\rho)^{2/3} H(\sigma)^2
\end{equation}
We obtain
\begin{align}
	G' &= c_1 e^{a/2} \\
	\frac{H''}{H} &= N_f
\end{align}
where $c_1$ is a constant. In the case where $a=0$ and $N_f$ is a constant, we can solve this and find
\begin{equation}
	k = (\rho + \rho_0)^2
\end{equation}
and
\begin{align}
	j &= (c_1 \rho + c_2)^{2/3} \cos (\sqrt{-N_f} \sigma + c_3)^2 &\qquad \text{if } N_f \leq 0 \\
	j &= (c_1 \rho + c_2)^{2/3} \cosh (\sqrt{N_f} \sigma + c_3)^2 &\qquad \text{if } N_f \geq 0
\end{align}
with $c_1, c_2$ and $c_3$ are integration constants. These provide analytic solutions to the equations of motion of type IIB supergravity with modified Bianchi identity. When looking at the dilaton behavior, we find
\begin{align}
	e^{2\Phi} &= \frac{3 (\rho + \rho_0)}{c_1 (c_2 + c_1 \rho)^{1/3} \cos (c_3 + \sqrt{-N_f} \sigma )^4} &\qquad \text{if } N_f \leq 0 \label{eq:dilNfneg} \\
	e^{2\Phi} &= \frac{3 (\rho + \rho_0)}{c_1 (c_2 + c_1 \rho)^{1/3} \cosh (c_3 + \sqrt{N_f} \sigma )^4} &\qquad \text{if } N_f \geq 0
\end{align}
When $N_f \leq 0$, in \eqref{eq:dilNfneg}, it is remarkable that there are singularities for $c_3 + \sqrt{-N_f} \sigma = \frac{\pi}{2}$ mod $(2\pi)$. Those singularities may be a sign of the presence of the smeared flavor branes.

Another possibility would be to try to have a smearing form independent of one of the radial coordinates, instead of just the function $N_{f1}$ as in the previous paragraph. For $\Omega$ to be independant of $\sigma$, we have to take
\begin{equation}
	N_{f1} = \frac{N(\rho)}{\sqrt{j}}
\end{equation}
Then \eqref{eq:j-equation} becomes
\begin{equation}
	0 = 2 j^2 \partial^2_{\rho} j + 2 e^a j \partial^2_{\sigma} j + j (\partial_{\rho} j)^2 - e^a (\partial_{\sigma} j)^2 - j^2 a' \partial_{\rho} j - 4 e^a N(\rho) j^{3/2}
\end{equation}
Taking here $N(\rho)$ to be constant, we get $N_{f2} = 0$ which suppresses one of the terms in the smearing form. Nevertheless, it is not obvious how to find a solution to the equation for $j$.

For $\Omega$ to be independent of $\rho$, one needs to impose $k$ to be a constant. Then
\begin{align}
	a(\rho) &= 2 a_1 \rho \\
	N_{f1} &= \frac{e^{-a_1 \rho}}{\sqrt{j}} N(\sigma)
\end{align}
where $a_1$ is a strictly positive constant.
We now have to solve:
\begin{equation}
  0 = 2 j^2 \partial^2_{\rho} j + 2 e^{2 a_1 \rho} j \partial^2_{\sigma} j + j (\partial_{\rho} j)^2 - e^{2 a_1 \rho} (\partial_{\sigma} j)^2 - 2 j^2 a_1 \partial_{\rho} j - 4 e^{a_1 \rho} N(\sigma) j^{3/2} 
\end{equation}
In the case where $N(\sigma) = N_f$ is a constant, the smearing form is independent of any radial dependence. In that case we can find asymptotic solutions, considering $\rho$ as the energy scale. One interesting fact is that it seems it is not possible to ignore the term involving $N_f$ in the IR, that is when $\rho$ goes to zero. In the IR $(\rho \rightarrow 0)$, we find that
\begin{align}
	j &= e^{2a_1 \rho/3} \left( \frac{3N_f}{a_1^2-a_1}+ c_1 e^{(1-a_1)\rho} \right)^{2/3}  &\qquad \text{if } a_1 \neq 1 \\
	j &= e^{2a_1 \rho/3} \left( 3 N_f \rho + c_2 e^{-\rho} \right)^{2/3} &\qquad \text{if } a_1 = 1
\end{align}
In the UV, we have two possibilities: we can decide that the term in $N_f$ is suppressed or plays a role. The two cases give
\begin{align}
	j &= c_3 e^{2 a_1 \rho /3} \sigma^2 &\qquad \text{if we neglect the term in } N_f \\
	j &= e^{-2a_1 \rho} \frac{N_f^2 \sigma^4}{4}
\end{align}

\subsubsection{Comments on the solution}

Firstly one can notice that none of the solutions presented in the
previous section goes to the solution found in \cite{Gauntlett:2001ur}
in the limit $N_{f1}, N_{f2}$ goes to zero, as expected from the dual gauge theory point of vue.

We are trying to find a solution that describes a stack of $N_c$ color
branes plus one or several stacks of smeared flavor branes. The number
of color branes is related to the Ramond-Ramond field $F_{(3)}$ through
\begin{equation}
	\int_{S^3} F_{(3)} = 2 \kappa^2_{10} T_5 N_c
\end{equation}
where $S^3$ is a three-sphere around the point where the color branes
are placed in the four-dimensional space transverse to their
world-volume. We were not able to find a constant when calculating the
previous integral for the solutions of the previous section. It means
that either we did not find the right transverse four-dimensional
space, or these results cannot have the usual interpretation of stacks
of branes.

This relates to the most prominent problem of the method presented in
this section. As we mentioned in footnote \ref{fn:integrability}, it
is necessary to verify the existence of a cycle wrapped by the
branes. As we explicitly avoided the issue of considering the
embedding smeared, one cannot be certain that the above solutions do
describe smeared branes. In simple cases when the smearing form does not
have a term along the radial direction of the space, each component in
the vielbein basis can usually be interpreted as the volume form of
the space orthogonal to the brane smeared. In the case studied above,
$\Omega$ has to have a term in $\td \rho$. So in comparison to
Klebanov-Witten, it seems that we are smearing massive flavor
branes. But we were not able to determine their embedding. However,
the form of $\Omega$ tells us it is not possible to smear massless
flavor branes in this background. Moreover, knowing the explicit
embedding of the flavor branes is not necessary to look at some
properties of the gauge theory dual.

\section{Conclusion}
\label{sec:conclusion}
In this paper, we applied generalized calibrations and $G$-structures to
address the problem of 
adding smeared flavor branes to a supergravity background. In doing
so, we made a first step towards a systematic
study of backgrounds with a large number of smeared flavor branes.
In section \ref{sec:geometry-of-smeared-branes}, we showed that the
smeared brane action of \cite{Koerber:2006hh} is equivalent to those
used previously in the literature on smeared flavor branes. This makes
the symmetry with the Wess-Zumino 
term apparent and the linearity in the smearing form $\Omega$
manifest. Furthermore we were able to link the complete brane action
to a conserved charge and impose strong constraints on
$\Omega$ by relating it to the calibration form. While the explicit
form of $\Omega$ depends on the embedding smeared, this allowed us to
explain various features of the examples in section
\ref{sec:three-examples}; in particular why the smearing has to
preserve certain symmetries, which again implies that it is often only
possible to smear several stacks of branes at once.

We exhibited the potential of our methods not only by studying known
examples, yet by also flavoring a background dual to a $d=2+1$,
$\mathcal{N}=2$ super Yang-Mills-like theory (See section
\ref{sec:gauge-string-duality-in-d=2+1}). Here we found several
solutions and some interesting features, notably the fact that it is
not possible to smear massless flavors -- a property which would be
nice to understand from the point of view of the dual gauge theory.

The formalism introduced in this paper unifies the treatment of
different possible embeddings for any single background, allowing for
a general study of the smearing procedure in a given background,
instead of the case by case methods previously used. Even if it
remains necessary to verify the existence of the cycles wrapped by the
branes, their knowledge is not necessary for the actual calculation.
However, as we have seen in the case of the $d=2+1$, $\mathcal{N}=2$
duality, backgrounds constructed without knowledge of the embeddings
might be very difficult to interpret.

It would be interesting to work in the future on the removal of some
limitations of the work we presented. Extensions of the results of
this paper to type IIA backgrounds, world-volume gauge fields or the
Kalb-Ramond field should be straightforward using the results of
\cite{Martucci:2005ht}. While we were able to impose
strong mathematical constraints onto the smearing form $\Omega$, we
were not able to link it to the physical interpretation of a brane
density. In other words, we are not providing a general way of knowing
from the smearing form and the ansatz for $N_f(\rho)$ what the
embeddings of the smeared flavor branes are. Even if such knowledge is
not required to study some aspects of the gauge theory dual, it would
give a better understanding of the way the duality is working. One
might also wonder how much one can learn about the various dual gauge
theories from the generic form of $\Omega$ prior to selecting one of
them by making an ansatz for $N_f(\rho)$.

Finally, it might turn out to be useful to apply further relations
between supersymmetry and geometry to the field of gauge/string
duality. Examples of this are given by generalized complex geometry or
the use of pure spinors.

\section*{Acknowledgements}
\label{sec:acknowledgements}
We would like to thank Carlos N\'u\~nez. We would also like to thank Prem Kumar
and Ioannis Papadimitriou for useful discussions and Daniel Are\'an,
Francesco Bigazzi, Aldo Cotrone, Dario Martelli, \'Angel Paredes and
Alfonso Ramallo for their comments on the manuscript. J.S.~is
supported by the German National Academic Foundation (Studienstiftung
des deutschen Volkes) and an STFC studentship.

\appendix

\section{A review of generalized calibrated geometry}
\label{sec:short-review-of-calibrations}

We shall give a short introduction to generalized calibrated
geometry \cite{Gutowski:1999tu} in relation to supersymmetric brane
embeddings. The discussion given ignores the case of world-volume
fluxes and follows that of the review \cite{Gauntlett:2003di}.

\paragraph{Calibration forms and supersymmetric brane embeddings}
\label{sec:calibration-forms-and-supersymmetric-brane-embeddings}

The standard method used when studying supersymmetric brane embeddings
is $\kappa$-sym\-me\-try \cite{Bergshoeff:1996tu}. A brane embedding
$X^M(\xi)$ is supersymmetric if it satisfies the equation
\begin{equation}
  \label{eq:kappa-symmetry-condition}
  \Gamma_\kappa \epsilon = \epsilon
\end{equation}
$\epsilon$ is a SUSY spinor of the background while $\Gamma_\kappa$ is
a linear map depending on the form of the embedding. For a D-brane of
type IIB string theory with world-volume gauge fields such that
$\mathcal{F} = 2\pi \alpha^\prime F - \mathcal{B} = 0$ it reduces to
\begin{equation}
  \label{eq:kappa-symmetry-matrix}
  \Gamma_\kappa = \frac{1}{(1+p)! \sqrt{-\hat{g}_{(p+1)}}}
  \epsilon^{\alpha_0 \dots \alpha_p} \left\{
    \begin{array}{ll}
      (\Gamma^{11})^{\frac{p-2}{2}}
      \gamma_{\alpha_0 \dots \alpha_p} & \textrm{(IIA)}\\
      \sigma_3^{\frac{p-3}{2}} \imath \sigma_2 \otimes
      \gamma_{\alpha_0 \dots \alpha_p} & \textrm{(IIB)}
    \end{array}
    \right.
\end{equation}
which is invariant under Weyl transformations and therefore valid in
string and Einstein frame. The definition uses the pull-back of the
space-time $\Gamma$-matrices onto the brane world-volume,
$\gamma_\alpha = \partial_\alpha X^M \Gamma_M$.

$\Gamma_\kappa$ is hermitian and squares to one. It follows that
\begin{equation}
  \label{eq:derivation-of-calibration-inequality}
  \begin{aligned}
    \epsilon^\dagger \frac{1-\Gamma_\kappa}{2} \epsilon
    &= \epsilon^\dagger \frac{1-\Gamma_\kappa}{2}
    \frac{1-\Gamma_\kappa}{2} \epsilon = \left\Vert
    \frac{1-\Gamma_\kappa}{2} \epsilon \right\Vert^2 \geq 0
  \end{aligned}
\end{equation}
Which implies that $\epsilon^\dagger \epsilon \geq \epsilon^\dagger
\Gamma_\kappa \epsilon$ with equality if and only if the embedding is
supersymmetric. Normalizing the spinor such that $\epsilon^\dagger
\epsilon = 1$ and using (\ref{eq:kappa-symmetry-matrix}), we may
rephrase this as
\begin{equation}
  \label{eq:calibration-form-inequality}
  \sqrt{-\hat{g}_{(p+1)}} \geq \frac{1}{(p+1)!} \epsilon^{\alpha_0 \dots
    \alpha_p} \left\{
    \begin{array}{ll}
      \epsilon^\dagger
        (\Gamma^{11})^{\frac{p-2}{2}}
        \gamma_{\alpha_0 \dots \alpha_p} \epsilon &
      \textrm{(IIA)}\\ 
      \epsilon^\dagger
        \sigma_3^{\frac{p-3}{2}} \imath \sigma_2 \otimes
        \gamma_{\alpha_0 \dots \alpha_p} \epsilon &
      \textrm{(IIB)}
    \end{array}
    \right.
\end{equation}
Equality holds if and only if the embedding is supersymmetric.
Now the right hand side of (\ref{eq:calibration-form-inequality})
may be written as the pull-back of a differential form defined in
space-time.
\begin{equation}
  \label{eq:definition-of-calibration-form}
  \hat{\phi} = \frac{1}{(p+1)!} e^{a_0 \dots a_p} \left\{
    \begin{array}{ll}
      \epsilon^\dagger (\Gamma^{11})^{\frac{p-2}{2}} \Gamma_{a_0 \dots
        a_p} \epsilon & \textrm{(IIA)} \\
      \epsilon^\dagger
      \sigma_3^{\frac{p-3}{2}} \imath
      \sigma_2 \otimes \Gamma_{a_0 \dots a_p} \epsilon & \textrm{(IIB)}
    \end{array} 
  \right.
\end{equation}
$\hat{\phi}$ is known as the calibration form. A criterion for supersymmetry
of an embedding that is alternative to
(\ref{eq:kappa-symmetry-condition}) is then given by the following
\begin{equation}
  \label{eq:SUSY-brane-in-terms-of-calibration}
  X^* \hat{\phi} = \sqrt{-\hat{g}_{(p+1)}} \td^{p+1}\xi
\end{equation}
that is, the pull-back of the calibration form onto the world-volume is
equal to the induced volume form.

One may obtain $\hat{\phi}$ directly from its definition
(\ref{eq:definition-of-calibration-form}) and the knowledge of the
projections imposed onto the SUSY spinors. We shall give an example of
this in appendix
\ref{sec:explicit-example-of-calculating-a-calibration-form}.

\paragraph{A more formal definition}
\label{sec:formal-definition-of-calibrations}

Formally one defines a calibration on a Riemannian manifold as a
$(p+1)$-form $\hat{\phi}$ satisfying
\begin{equation}
  \label{eq:formal-definition-of-calibration}
  \begin{aligned}
    \td \hat{\phi} &= 0 &
    &\qquad &
    \hat{\phi} \vert_{\xi^{p+1}} &\leq \eta_{(p+1)} \vert_{\xi^{p+1}}
  \end{aligned}
\end{equation}
Here $\xi^p$ is a set of vectors specifying a tangent $(p+1)$-plane to a
$(p+1)$-cycle $\Sigma_{p+1}$ while $\eta_{(p+1)} =
\sqrt{-\hat{g}_{(p+1)}} \td^{p+1} \xi$ is the volume form induced onto
that cycle. The cycle $\Sigma_{p+1}$ is calibrated if the above bound
is saturated, i.e.~if $\hat{\phi} \vert_{\xi^{p+1}} = \eta_{(p+1)} \vert_{\xi^{p+1}}$.

As we have seen above (\ref{eq:SUSY-brane-in-terms-of-calibration}),
$\kappa$-symmetric brane embeddings satisfy the volume bound, which
can be thought of as a BPS-bound. In this and in the next paragraph we
shall turn to the issue of the closure of
(\ref{eq:definition-of-calibration-form}). For a background without
fluxes, the issue is rather easily resolved. From the gravitino
variation
\begin{equation}
  \delta_\epsilon \psi_M = D_M \epsilon = 0
\end{equation}
it follows that the SUSY spinor $\epsilon$ is covariantly constant. As
the covariant derivative of both the vielbein and the tangent-space
$\Gamma$-matrices does also vanish it follows that
\begin{equation}
  \begin{aligned}
    \td \hat{\phi} &= \nabla \wedge \hat{\phi} = 0
  \end{aligned}
\end{equation}
$\nabla \wedge \hat{\phi}$ is to be thought of as a formal expression. The
wedge product antisymmetrizes over the relevant indices and, as the
Levi-Civita connection is symmetric in two of its indices, it follows
that the first equality holds. As all the ingredients of
(\ref{eq:definition-of-calibration-form}) are covariantly constant, it
follows that the exterior derivative is closed.

There is a nice interpretation of the closure of the calibration
form. Let us assume that we deform the calibrated cycle $\Sigma_{p+1}$ to
$\Sigma_{p+1}^\prime$. The two cycles differ by a boundary $\Sigma_{p+1} -
\Sigma_{p+1}^\prime = \delta \Xi_{p+2}$. More formally we would not
consider $\Sigma_{p+1}^\prime$ as a deformation, yet as a cycle within the
homology class defined by $\Sigma_{p+1}$. We use Stokes theorem to
establish
\begin{equation}
  \label{eq:minimal-volume-relation-for-calibrations}
  \textrm{Vol}(\Sigma_{p+1}) = \int_{\Sigma_{p+1}} \hat{\phi} =
  \int_{\Xi_{p+2}} \td \hat{\phi} + \int_{\Sigma_{p+1}^\prime} \hat{\phi} =
  \int_{\Sigma_{p+1}^\prime} \hat{\phi} \leq
  \textrm{Vol}(\Sigma_{p+1}^\prime)
\end{equation}
The final inequality uses (\ref{eq:calibration-form-inequality}).
It follows that the calibrated cycle $\Sigma_{p+1}$ is a minimal
volume cycle. This matches nicely with our experience from string
theory. In the absence of fluxes branes wrap minimal volume cycles.

\paragraph{Generalized calibrations}
\label{sec:generalized-calibrations}

The $\kappa$-symmetry matrix (\ref{eq:kappa-symmetry-matrix}) does not
change in the presence of Ramond-Ramond background fields and thus
neither does the definition of the calibration form or the
supersymmetry condition
(\ref{eq:SUSY-brane-in-terms-of-calibration}). Background fluxes
however deform branes such that they do not longer wrap
minimal volume cycles. For a background with fluxes we do therefore
not expect the calibration form
(\ref{eq:definition-of-calibration-form}) to be closed. Rather, it's
exterior differential should be related to the flux. Indeed, in all
the examples studied in section \ref{sec:three-examples} the
calibration satisfied
\begin{equation}
  \label{eq:exterior-differential-of-generalized-calibration}
  \td (e^{\frac{p-3}{4}\Phi} \hat{\phi}) = F_{(p+2)}
\end{equation}
In this case, one speaks of a generalized calibration, a concept which
was first introduced in \cite{Gutowski:1999tu}.

There are several ways to prove
(\ref{eq:exterior-differential-of-generalized-calibration}). For all
the examples of \ref{sec:three-examples}, the equality held after we
imposed the BPS equations, so it should be no surprise that
(\ref{eq:exterior-differential-of-generalized-calibration}) is
intimately linked to the supersymmetry of the background. The original
proof \cite{Gutowski:1999tu} showed that the expression
$(e^{\frac{p-3}{4}\Phi} \hat{\phi} - C_{(p+1)})$ appears as the central
charge of a supersymmetry algebra and must therefore be topological
and thus exact. It is also possible to verify
(\ref{eq:exterior-differential-of-generalized-calibration}) in terms
of the dilatino and gravitino supersymmetry transformations.

Before doing so however, we shall take a look at the appropriate
generalization of
(\ref{eq:minimal-volume-relation-for-calibrations}). To do so we shall
assume that both the brane and the background fields are static. It
follows that the energy of the system is proportional to its action
-- with the proportionality constant being infinity. Moreover, minimum
energy configurations will therefore minimize the brane action. Let
$\Sigma_{p+1}$ be the supersymmetric cycle wrapped by the brane and
$\Sigma_{p+1}^\prime = \Sigma_{p+1} + \delta \Xi_{(p+2)}$ a
deformation. Then (setting $T_p = 1$)
\begin{equation}
  \label{eq:energy-inequality-for-generalized calibrations}
  \begin{aligned}
    \Delta E &\propto S_{\Sigma_{p+1}^\prime} - S_{\Sigma_{p+1}} \\
    &= \int_{\Sigma_{p+1}^\prime} (e^{\frac{p-3}{4}\Phi} \eta -
    C_{(p+1)}) - \int_{\Sigma_{p+1}} (e^{\frac{p-3}{4}\Phi} \hat{\phi} -
    C_{(p+1)}) \\
    &\geq \int_{\delta \Xi_{p+2}} (e^{\frac{p-3}{4}\Phi} \hat{\phi} -
    C_{(p+1)}) \\
    &= \int_{\Xi_{p+2}} \td (e^{\frac{p-3}{4}\Phi} \hat{\phi} -
    C_{(p+1)}) = 0
  \end{aligned}
\end{equation}
The inequality in the second line used again
(\ref{eq:calibration-form-inequality}). It follows that
supersymmetric, static embeddings are minimum energy configurations.

\section{Finding the calibration form -- an explicit example}
\label{sec:explicit-example-of-calculating-a-calibration-form}

As an example we will calculate the calibration form for the theory of
section \ref{sec:gauge-string-duality-in-d=2+1}. Apart from the
definition (\ref{eq:definition-of-calibration-form}) we will need the
projections imposed on the background SUSY spinors. To simplify things
we perform a change of basis on the spinors taking $\sigma_1 \mapsto
\sigma_3$. As a result of this transformations, the two Majorana-Weyl
spinors in $\zeta = \left( \begin{smallmatrix} \zeta^- \\
    \zeta^+ \end{smallmatrix} \right)$ decouple
\begin{equation}
  \begin{aligned}
    \Gamma^{1256} \zeta^\mp &= \zeta^\mp &
    \Gamma^{1346} \zeta^\mp &= \zeta^\mp &
    \Gamma^{4567} \zeta^\mp &= \pm \zeta^\mp
  \end{aligned}
\end{equation}
We will also need the fact that IIB supergravity is chiral, with the chirality
chosen such that
\begin{equation}
  \Gamma_{11} \zeta^\mp = \Gamma_{1 2 3 \dots 8 9 0} \zeta^\mp = -\zeta^\mp
\end{equation}
Note that our change of basis does also affect the definition of the
calibration form (\ref{eq:definition-of-calibration-form}) -- we
obtain two calibration forms, $\hat{\phi}^\mp$. Note also that we will work
in flat indices.

Before looking at the most generic case, we shall look at a few examples of
how to calculate components of $\hat{\phi}^\mp$
\begin{equation}
  \begin{aligned}
    \hat{\phi}^\mp_{089123} &= \zeta^{\mp T} \Gamma_{089123} \zeta^\mp
    = \zeta^{\mp T} \Gamma_{4567} \zeta^\mp = \pm 1 \\
    \hat{\phi}^\mp_{089145} &= \zeta^{\mp T} \Gamma_{2367} \zeta^\mp =
    \zeta^{\mp T} \Gamma_{4567} \zeta^\mp = \pm 1 \\
    \hat{\phi}^\mp_{089167} &= \zeta^{\mp T} \Gamma_{2345} \zeta^\mp = 1
  \end{aligned}
\end{equation}
These examples show nicely that the two forms disagree on those cycles
making use of the $\Gamma^{4567}$ projection. As the two forms need to
disagree by an overall sign for a cycle to be supersymmetric, it
follows that cycles involving the $7$ direction cannot be
supersymmetric. One can arrive at the same result directly from the
$\kappa$-symmetry condition.

The more difficult step is to show why components such
as
\begin{equation}
  \begin{aligned}
    \hat{\phi}^\mp_{089567} &= \zeta^{\mp T} \Gamma_{1234} \zeta^\mp =
    \zeta^{\mp T} \Gamma_{26} \zeta^\mp
  \end{aligned}
\end{equation}
vanish.
Starting from the projections
\begin{equation}\label{eq:calibration_form_spinor_projections}
  \begin{aligned}
      \frac{1 - \Gamma^{1346}}{2} \zeta^\mp &= 0 &
      \frac{1 - \Gamma^{1256}}{2} \zeta^\mp &= 0 &
      \frac{1 \mp \Gamma^{4567}}{2} \zeta^\mp &= 0
  \end{aligned}
\end{equation}
we define orthogonal projectors
\begin{equation}\label{eq:calibration_form_orthogonal_spinor_projections}
  \begin{aligned}
    &\frac{1 + \Gamma^{1346}}{2} &
    &\frac{1 + \Gamma^{1256}}{2} &
    &\frac{1 \pm \Gamma^{4567}}{2}
  \end{aligned}
\end{equation}
which may be used to project an arbitrary spinor $\psi$ onto the
subspace of spinors satisfying
(\ref{eq:calibration_form_spinor_projections}) because
\begin{equation}
    \left( \frac{1 - \Gamma^{1346}}{2} \right) \left( \frac{1 +
        \Gamma^{1346}}{2} \right) \psi= 0
\end{equation}
independently of the choice of $\psi$. This is simply the
defining property of orthogonal projections.
Note that $\zeta$ may be assumed to be invariant under the
orthogonal projections, as it satisfies
(\ref{eq:calibration_form_spinor_projections}). Applying this to the
question of $\hat{\phi}_{089567}^\mp$,
\begin{equation}
  \begin{aligned}
    \hat{\phi}^\mp_{089567} &= \zeta^{\mp T} \Gamma_{26} \zeta^\mp \\
    &= \frac{1}{4} \left\lbrack (1+\Gamma^{1346}) \zeta
    \right\rbrack^T \Gamma^{26} (1+\Gamma^{1346}) \zeta \\
    &= \frac{1}{4} \zeta^{\mp T} (\Gamma^{26} - \Gamma^{1234} +
    \Gamma^{1234} - \Gamma^{26}) \zeta^\mp \\
    &= 0
  \end{aligned}
\end{equation}
Note however that it does not appear to be obvious which of the
projections (\ref{eq:calibration_form_orthogonal_spinor_projections})
one has to choose to show that a particular component of $\hat{\phi}$
vanishes. To give an example of this, let's look at
\begin{equation}
  \begin{aligned}
    \hat{\phi}^\mp_{089124} &= \pm \zeta^{\mp T} \Gamma_{34} \zeta^\mp \\
    &= \pm \frac{1}{4} \zeta^{\mp T} (1+\Gamma^{1346}) \Gamma^{34}
    (1+\Gamma^{1346}) \zeta^\mp \\
    &= \pm \frac{1}{4} \zeta^{\mp T} (\Gamma^{34} - \Gamma^{16} -
    \Gamma^{16} + \Gamma^{34}) \zeta^\mp \\
    \hat{\phi}^\mp_{089124} &= \pm \frac{1}{4} \zeta^{\mp T} (1 \pm
    \Gamma^{4567}) 
    \Gamma^{34} (1 \pm \Gamma^{4567}) \zeta^\mp \\
    &= \frac{1}{4} \zeta^{\mp T} (\pm \Gamma^{34} -
    \Gamma^{3567})(1 \pm \Gamma^{4567}) \zeta^\mp \\
    &= \frac{1}{4} \zeta^{\mp T} (\pm \Gamma^{34} - \Gamma^{3567} +
    \Gamma^{3567} \mp \Gamma^{34}) \zeta^\mp 
    = 0
  \end{aligned}
\end{equation}
Let's try to look at a generic case. There is no summation in the
following. Instead the indices $(a, b, c, d, e, f, g) \in \{ 1, \dots,
7 \}$ are all independent and mutually non-equal, $a \neq b, a \neq c,
\dots, f \neq g$.
\begin{equation}
  \label{eq:calibration_form_from_spinors_super-equation}
  \begin{aligned}
    \pm 4 \hat{\phi}^\mp_{089abc} &= \zeta^{\mp T} (1 \pm \Gamma^{cdef})
    \Gamma^{defg} (1 \pm \Gamma^{cdef}) \zeta^\mp \\
    &= \zeta^{\mp T} (\Gamma^{defg} \mp \Gamma^{cg})(1 \pm \Gamma^{cdef})
    \zeta^\mp \\
    &= \zeta^{\mp T} (\Gamma^{defg} \mp \Gamma^{cg} \pm \Gamma^{cg} -
    \Gamma^{defg}) \zeta^\mp = 0
  \end{aligned}
\end{equation}
In the first line we used the chirality matrix $\Gamma_{11}$ to change
$\Gamma_{089abc}$ into $\Gamma^{defg}$. In the process we might have
picked up an overall minus sign, which we moved together with the
factor $4$ to the left hand side. In the projection matrices we have
$\Gamma$-matrices assumed to be of the form $\Gamma^{cdef}$. Here there
is again a sign ambiguity, as we have moved $c$ to the left and as the
projection might involving $\Gamma^7$. Note
that we have the same sign in both parentheses, so in the following
lines we will always have either the upper signs or the lower signs,
never a mixture of the two -- which is why $\pm \mp = -$ in the second
to last equality. Similarly we shall now take a look at
\begin{equation}
  \label{eq:calibration_form_from_spinors_super-equation-2}
  \begin{aligned}
    \pm 4 \hat{\phi}^\mp_{089abc} &= \zeta^{\mp T} (1 \pm \Gamma^{abcd})
    \Gamma^{defg} (1 \pm \Gamma^{abcd}) \zeta^\mp \\
    &= \zeta^{\mp T} (\Gamma^{defg} \pm \Gamma^{abcefg} \mp
    \Gamma^{abcefg} - \Gamma^{defg}) \zeta^\mp = 0
  \end{aligned}
\end{equation}

Equation (\ref{eq:calibration_form_from_spinors_super-equation}) is a
very potent result. It follows immediately that
\begin{equation}
  \begin{aligned}
    \hat{\phi}_{124} &= 0 &
    \hat{\phi}_{125} &= 0 &
    \hat{\phi}_{126} &= 0 &
    \hat{\phi}_{127} &= 0 &
    \hat{\phi}_{134} &= 0 &
    \hat{\phi}_{135} &= 0 \\
    \hat{\phi}_{136} &= 0 &
    \hat{\phi}_{137} &= 0 &
    \hat{\phi}_{147} &= 0 &
    \hat{\phi}_{157} &= 0 &
    \hat{\phi}_{234} &= 0 &
    \hat{\phi}_{235} &= 0 \\
    \hat{\phi}_{236} &= 0 &
    \hat{\phi}_{237} &= 0 &
    \hat{\phi}_{245} &= 0 &
    \hat{\phi}_{247} &= 0 &
    \hat{\phi}_{256} &= 0 &
    \hat{\phi}_{267} &= 0 \\
    \hat{\phi}_{345} &= 0 &
    \hat{\phi}_{346} &= 0 &
    \hat{\phi}_{357} &= 0 &
    \hat{\phi}_{367} &= 0 &
    \hat{\phi}_{457} &= 0 &
    \hat{\phi}_{467} &= 0
  \end{aligned}
\end{equation}
Similarly we gather from
(\ref{eq:calibration_form_from_spinors_super-equation-2})
\begin{equation}
  \begin{aligned}
    \hat{\phi}_{146} &= 0 &
    \hat{\phi}_{156} &= 0 &
    \hat{\phi}_{456} &= 0
  \end{aligned}
\end{equation}

All these things considered we are able to reproduce the two
calibration forms exhibited in \cite{Gauntlett:2001ur},
\begin{equation}
  \begin{aligned}
    \hat{\phi}^- &= e^{089} \wedge \left( e^{123} + e^{145} - e^{167} +
      e^{246} + e^{257} + e^{347} - e^{356} \right) \\ 
    \hat{\phi}^+ &= e^{089} \wedge \left( - e^{123} - e^{145} - e^{167} -
      e^{246} + e^{257} + e^{347} + e^{356} \right)
  \end{aligned}
\end{equation}
There is a second result following immediately from equations
(\ref{eq:calibration_form_from_spinors_super-equation}) and
(\ref{eq:calibration_form_from_spinors_super-equation-2}). For the
SUSY projections not to be mutually exclusive they have to have
pairwise two indices in common. Note that this may be easily
generalized to arbitrary dimensions. In general one finds that if the
SUSY projections take the form of antisymmetrized Gamma matrices with
four indices, $\Gamma^{abcd} \zeta = \zeta$, different projections
have to have an even number of indices in common (zero or two; four
means that the projections are equal) in order to be
compatible. Compatible means that this requirement is necessary for a
spinor $\zeta$ satisfying all projections to exist. This result simply
requires the properties of the Dirac algebra.

\bibliographystyle{plain}


\begin{thebibliography}{99}

\bibitem{Maldacena:1997re}
 J.~M.~Maldacena,
 Adv.\ Theor.\ Math.\ Phys.\  {\bf 2}, 231 (1998)
 [Int.\ J.\ Theor.\ Phys.\  {\bf 38}, 1113 (1999)]
 [arXiv:hep-th/9711200].

\bibitem{Witten:1998qj}
 E.~Witten,
 Adv.\ Theor.\ Math.\ Phys.\  {\bf 2}, 253 (1998)
 [arXiv:hep-th/9802150].

\bibitem{Klebanov:1998hh}
 I.~R.~Klebanov and E.~Witten,
 Nucl.\ Phys.\  B {\bf 536}, 199 (1998)
 [arXiv:hep-th/9807080].

\bibitem{Klebanov:2000hb}
 I.~R.~Klebanov and M.~J.~Strassler,
 JHEP {\bf 0008}, 052 (2000)
 [arXiv:hep-th/0007191].

\bibitem{Klebanov:2000nc}
 I.~R.~Klebanov and A.~A.~Tseytlin,
 Nucl.\ Phys.\  B {\bf 578}, 123 (2000)
 [arXiv:hep-th/0002159].

\bibitem{Witten:1998zw}
 E.~Witten,
 Adv.\ Theor.\ Math.\ Phys.\  {\bf 2}, 505 (1998)
 [arXiv:hep-th/9803131].

\bibitem{Maldacena:2000yy}
 J.~M.~Maldacena and C.~Nunez,
 Phys.\ Rev.\ Lett.\  {\bf 86}, 588 (2001)
 [arXiv:hep-th/0008001].

\bibitem{Gomis:2001aa}
 J.~Gomis and J.~G.~Russo,
 JHEP {\bf 0110}, 028 (2001)
 [arXiv:hep-th/0109177].

\bibitem{Gauntlett:2001ps}
 J.~P.~Gauntlett, N.~Kim, D.~Martelli and D.~Waldram,
 Phys.\ Rev.\  D {\bf 64}, 106008 (2001)
 [arXiv:hep-th/0106117].

\bibitem{Bigazzi:2001aj}
 F.~Bigazzi, A.~L.~Cotrone and A.~Zaffaroni,
 Phys.\ Lett.\  B {\bf 519}, 269 (2001)
 [arXiv:hep-th/0106160].

\bibitem{Karch:2002sh}
 A.~Karch and E.~Katz,
 JHEP {\bf 0206}, 043 (2002)
 [arXiv:hep-th/0205236].

\bibitem{Klebanov:2004ya}
 I.~R.~Klebanov and J.~M.~Maldacena,
 Int.\ J.\ Mod.\ Phys.\  A {\bf 19}, 5003 (2004)
 [arXiv:hep-th/0409133].

\bibitem{Burrington:2004id}
 B.~A.~Burrington, J.~T.~Liu, L.~A.~Pando Zayas and D.~Vaman,
 JHEP {\bf 0502}, 022 (2005)
 [arXiv:hep-th/0406207].

\bibitem{Kirsch:2005uy}
 I.~Kirsch and D.~Vaman,
 Phys.\ Rev.\  D {\bf 72}, 026007 (2005)
 [arXiv:hep-th/0505164].

\bibitem{Bigazzi:2005md}
 F.~Bigazzi, R.~Casero, A.~L.~Cotrone, E.~Kiritsis and A.~Paredes,
 JHEP {\bf 0510}, 012 (2005)
 [arXiv:hep-th/0505140].

\bibitem{Casero:2006pt}
 R.~Casero, C.~Nunez and A.~Paredes,
 Phys.\ Rev.\  D {\bf 73}, 086005 (2006)
 [arXiv:hep-th/0602027].

\bibitem{Casero:2007jj}
 R.~Casero, C.~Nunez and A.~Paredes,
 Phys.\ Rev.\  D {\bf 77}, 046003 (2008)
 [arXiv:0709.3421 [hep-th]].

\bibitem{HoyosBadajoz:2008fw}
 C.~Hoyos-Badajoz, C.~Nunez and I.~Papadimitriou,
 Phys.\ Rev.\  D {\bf 78}, 086005 (2008)
 [arXiv:0807.3039 [hep-th]].

\bibitem{Benini:2006hh}
 F.~Benini, F.~Canoura, S.~Cremonesi, C.~Nunez and A.~V.~Ramallo,
 JHEP {\bf 0702}, 090 (2007)
 [arXiv:hep-th/0612118].

\bibitem{Caceres:2007mu}
 E.~Caceres, R.~Flauger, M.~Ihl and T.~Wrase,
 JHEP {\bf 0803}, 020 (2008)
 [arXiv:0711.4878 [hep-th]].

\bibitem{Benini:2007gx}
 F.~Benini, F.~Canoura, S.~Cremonesi, C.~Nunez and A.~V.~Ramallo,
 JHEP {\bf 0709}, 109 (2007)
 [arXiv:0706.1238 [hep-th]].

\bibitem{Casero:2007pz}
 R.~Casero and A.~Paredes,
 Fortsch.\ Phys.\  {\bf 55}, 678 (2007)
 [arXiv:hep-th/0701059].

\bibitem{Benini:2007kg}
 F.~Benini,
 JHEP {\bf 0810}, 051 (2008)
 [arXiv:0710.0374 [hep-th]].

\bibitem{Bigazzi:2008zt}
 F.~Bigazzi, A.~L.~Cotrone and A.~Paredes,
 JHEP {\bf 0809}, 048 (2008)
 [arXiv:0807.0298 [hep-th]].

\bibitem{Canoura:2008at}
 F.~Canoura, P.~Merlatti and A.~V.~Ramallo,
 JHEP {\bf 0805}, 011 (2008)
 [arXiv:0803.1475 [hep-th]].

\bibitem{Arean:2008az}
 D.~Arean, P.~Merlatti, C.~Nunez and A.~V.~Ramallo,
 arXiv:0810.1053 [hep-th].

\bibitem{Paredes:2006wb}
 A.~Paredes,
 JHEP {\bf 0612}, 032 (2006)
 [arXiv:hep-th/0610270].

\bibitem{Zeng:2007ta}
 D.~f.~Zeng,
 arXiv:0708.3814 [hep-th].

\bibitem{Bigazzi:2008gd}
 F.~Bigazzi, A.~L.~Cotrone, C.~Nunez and A.~Paredes,
 arXiv:0806.1741 [hep-th].

\bibitem{Bigazzi:2008ie}
 F.~Bigazzi, A.~L.~Cotrone, A.~Paredes and A.~Ramallo,
 arXiv:0810.5220 [hep-th].

\bibitem{Gutowski:1999tu}
 J.~Gutowski, G.~Papadopoulos and P.~K.~Townsend,
 Phys.\ Rev.\  D {\bf 60}, 106006 (1999)
 [arXiv:hep-th/9905156].

\bibitem{Koerber:2005qi}
  P.~Koerber,
  JHEP {\bf 0508}, 099 (2005)
  [arXiv:hep-th/0506154].

\bibitem{Martucci:2005ht}
  L.~Martucci and P.~Smyth,
  JHEP {\bf 0511}, 048 (2005)
  [arXiv:hep-th/0507099].

\bibitem{Koerber:2006hh}
  P.~Koerber and L.~Martucci,
  JHEP {\bf 0612}, 062 (2006)
  [arXiv:hep-th/0610044].

\bibitem{Gauntlett:2003cy}
 J.~P.~Gauntlett, D.~Martelli and D.~Waldram,
 Phys.\ Rev.\  D {\bf 69}, 086002 (2004)
 [arXiv:hep-th/0302158].

\bibitem{Gauntlett:2001ur}
 J.~P.~Gauntlett, N.~Kim, D.~Martelli and D.~Waldram,
 JHEP {\bf 0111}, 018 (2001)
 [arXiv:hep-th/0110034].

\bibitem{Nunez:2003cf}
 C.~Nunez, A.~Paredes and A.~V.~Ramallo,
 JHEP {\bf 0312}, 024 (2003)
 [arXiv:hep-th/0311201].

\bibitem{Koerber:2007hd}
 P.~Koerber and D.~Tsimpis,
 JHEP {\bf 0708}, 082 (2007)
 [arXiv:0706.1244 [hep-th]].

\bibitem{Arean:2004mm}
 D.~Arean, D.~E.~Crooks and A.~V.~Ramallo,
 JHEP {\bf 0411}, 035 (2004)
 [arXiv:hep-th/0408210].

\bibitem{DiVecchia:2002ks}
  P.~Di Vecchia, A.~Lerda and P.~Merlatti,
  Nucl.\ Phys.\  B {\bf 646}, 43 (2002)
  [arXiv:hep-th/0205204].

\bibitem{Gauntlett:2003di}
 J.~P.~Gauntlett,
 arXiv:hep-th/0305074.

\bibitem{Bergshoeff:1996tu}
 E.~Bergshoeff and P.~K.~Townsend,
 Nucl.\ Phys.\  B {\bf 490}, 145 (1997)
 [arXiv:hep-th/9611173].
\end{thebibliography}

\end{document}